\documentclass[11pt, oneside]{amsart}
\pdfoutput=1

\usepackage{amsmath}
\usepackage{amssymb}

\usepackage{color}
\usepackage{dcolumn}
\usepackage{float}
\usepackage{graphicx}
\usepackage[utf8]{inputenc}
\usepackage[T1]{fontenc}
\usepackage{lmodern}
\usepackage{multirow}
\usepackage{rotating}
\usepackage{subfigure}
\usepackage{psfrag}
\usepackage{tabularx}
\usepackage[hyphens]{url}
\usepackage{wrapfig}
\usepackage{longtable}
\usepackage{verbatim}
\usepackage{booktabs,multicol}

\usepackage{footnote}
\makesavenoteenv{tabular}
\makesavenoteenv{table}

\usepackage{enumitem}
\setlist{leftmargin=7mm}

\usepackage[bookmarks, bookmarksopen, bookmarksnumbered]{hyperref}
\usepackage[all]{hypcap}
\definecolor{orange}{rgb}{1.0,0.3,0.0}
\definecolor{violet}{rgb}{0.75,0,1}
\definecolor{darkgreen}{rgb}{0,0.6,0}
\definecolor{cyan}{rgb}{0.2,0.7,0.7}
\definecolor{blueish}{rgb}{0.2,0.2,0.8}
\definecolor{darkblue}{rgb}{0.1,0.1,0.9}

\newcommand{\note}[1]{ {\textcolor{blueish}    { ***Note:      #1 }}}

\usepackage[top=2.5cm, bottom=2.5cm, left=2.5cm, right=2.5cm]{geometry}

\sloppypar

\begin{document}

\title[]{Report on the Third Workshop on Sustainable Software for Science: Practice and Experiences (WSSSPE3)}

\author{Daniel S.\ Katz$^{(1)}$,
Sou-Cheng T.\ Choi$^{(2)}$,
Kyle E.\ Niemeyer$^{(3)}$,
James Hetherington$^{(4)}$,
Frank~L\"{o}ffler$^{(5)}$,
Dan Gunter$^{(6)}$,
Ray Idaszak$^{(7)}$,
Steven R.\ Brandt$^{(5)}$,
Mark A.\ Miller$^{(8)}$,
Sandra Gesing$^{(9)}$,
Nick D.\ Jones$^{(10)}$,
Nic Weber$^{(11)}$,
Suresh Marru$^{(12)}$,
Gabrielle Allen$^{(13)}$,
Birgit~Penzenstadler$^{(14)}$,
Colin C.\ Venters$^{(15)}$,
Ethan Davis$^{(16)}$,
Lorraine Hwang$^{(17)}$,
Ilian~Todorov$^{(18)}$,
Abani Patra$^{(19)}$,
Miguel de Val-Borro$^{(20)}$
}

\thanks{{}$^{(1)}$ \hspace{-1ex}Computation Institute, 
University of Chicago \& Argonne National Laboratory, Chicago, IL, USA; d.katz@ieee.org}
\thanks{{}$^{(2)}$ NORC at the University of Chicago and Illinois Institute of Technology, Chicago, IL, USA; sctchoi@uchicago.edu}
\thanks{{}$^{(3)}$ School of Mechanical, Industrial, and Manufacturing Engineering, 
Oregon State University, Corvallis, OR, USA; kyle.niemeyer@oregonstate.edu}
\thanks{{}$^{(4)}$ Research Software Development Group, University College London, UK; j.hetherington@ucl.ac.uk}
\thanks{{}$^{(5)}$ \hspace{-1.5ex}Center for Computation \& Technology, Louisiana State University, Baton Rouge, LA, USA; \{knarf,sbrandt\}@cct.lsu.edu}
\thanks{{}$^{(6)}$ Lawrence Berkeley National Laboratory, Berkeley, USA; dkgunter@lbl.gov}
\thanks{{}$^{(7)}$ RENCI, University of North Carolina at Chapel Hill, Chapel Hill, NC, USA; rayi@renci.org}
\thanks{{}$^{(8)}$ University of California, San Diego, CA, USA; mmiller@sdsc.edu}
\thanks{{}$^{(9)}$ Center for Research Computing, University of Notre Dame, Notre Dame, IN, USA; sandra.gesing@nd.edu}
\thanks{{}$^{(10)}$ New Zealand eScience Infrastructure (NeSI), University of Auckland, Auckland, NZ; nick.jones@nesi.org.nz}
\thanks{{}$^{(11)}$ University of Washington, Seattle, WA, USA; nmweber@uw.edu}
\thanks{{}$^{(12)}$ Indiana University, Bloomington, IN, USA; smarru@iu.edu}
\thanks{{}$^{(13)}$ National Center for Supercomputing Applications, University of Illinois at Urbana-Champaign, Urbana, IL, USA; gdallen@illinois.edu}
\thanks{{}$^{(14)}$ College of Computer Engineering \& Computer Science, California State University, Long Beach, CA, USA; birgit.penzenstadler@csulb.edu}
\thanks{{}$^{(15)}$ School of Computing and Engineering, University of Huddersfield, Huddersfield, UK; c.venters@hud.ac.uk}
\thanks{{}$^{(16)}$ UCAR Unidata, Boulder, CO, USA; edavis@ucar.edu}
\thanks{{}$^{(17)}$  University of California, Davis, CA, USA; ljhwang@ucdavis.edu}
\thanks{{}$^{(18)}$ Science \& Technology Facilities Council, UK; ilian.todorov@stfc.ac.uk}
\thanks{{}$^{(19)}$ Mechanical and Aerospace Engineering, University at Buffalo, Buffalo, NY, USA; abani@buffalo.edu}
\thanks{{}$^{(20)}$ Department of Astrophysical Sciences, Princeton University, Princeton, NJ, USA; valborro@princeton.edu}

\begin{abstract}
This report records and discusses the Third Workshop on Sustainable
Software for Science: Practice and Experiences (WSSSPE3). The report includes a
description of the keynote presentation of the workshop, which served as an overview of sustainable
scientific software. It also summarizes a set of lightning
talks in which speakers highlighted to-the-point lessons and challenges
pertaining to sustaining scientific software. The final and main contribution of the report is a summary of the
discussions, future steps, and future organization for a set of self-organized
working groups on topics including developing pathways to funding scientific
software; constructing useful common metrics for crediting software
stakeholders; identifying principles for sustainable software engineering
design; reaching out to research software organizations around the world; and
building communities for software sustainability. For each group, we include a
point of contact and a landing page that can be used by those who want to join
that group's future activities. The main challenge left by the workshop is to
see if the groups will execute these activities that they have scheduled, and
how the WSSSPE community can encourage this to happen.

\end{abstract}

\maketitle
\newpage

\section{Introduction} \label{sec:intro}

The Third Workshop on Sustainable Software for Science: Practice and Experiences
(WSSSPE3)\footnote{\url{http://wssspe.researchcomputing.org.uk/wssspe3/}} was
held on 28--29 September 2015 in Boulder, Colorado, USA. Previous events in the
WSSSPE series are
WSSSPE1\footnote{\url{http://wssspe.researchcomputing.org.uk/wssspe1/}}~\cite{WSSSPE1-pre-report,WSSSPE1},
held in conjunction with SC13;
WSSSPE1.1\footnote{\url{http://wssspe.researchcomputing.org.uk/wssspe1-1/}}, a
focused workshop organized jointly with the SciPy
conference\footnote{\url{https://conference.scipy.org/scipy2014/participate/wssspe/}};
WSSSPE2\footnote{\url{http://wssspe.researchcomputing.org.uk/wssspe2/}}~\cite{WSSSPE2-pre-report,WSSSPE2},
held in conjunction with SC14; and
WSSSPE2.1\footnote{\url{http://wssspe.researchcomputing.org.uk/wssspe2-1/}}, a
focused workshop organized again jointly with
SciPy\footnote{\url{http://scipy2015.scipy.org/ehome/115969/286469/}}.

Progress in scientific research is dependent on the quality and accessibility of
software at all levels. Hence it is critical to address challenges related to
development, deployment, maintenance, and overall sustainability of reusable
software as well as education around software practices. These challenges can be
technological, policy based, organizational, and educational; and are of
interest to developers (the software community), users (science disciplines),
software-engineering researchers, and researchers studying the conduct of
science (science of team science, science of organizations, science of science
and innovation policy, and social science communities). The WSSSPE1 workshop
engaged a broad scientific community to identify challenges and best practices
in areas of interest to creating sustainable scientific software. WSSSPE2
invited the community to propose and discuss specific mechanisms to move towards
an imagined future for software development and usage in science and
engineering. But WSSSPE2 did not have a good way to enact those mechanisms, or to
encourage the attendees to follow through on their intentions.

The WSSSPE3 workshop included multiple mechanisms for participation and
encouraged team building around solutions. WSSSPE3 strongly encouraged participation
of early-career scientists, postdoctoral researchers, graduate students,  
early-stage researchers, and those from underrepresented groups,
with funds provided to the conference organizers by the Moore Foundation, the
National Science Foundation (NSF), and the Software Sustainability Institute (SSI) to
support the travel of potential participants who would not otherwise be able to
attend the workshop. These
funds allowed 16 additional people to attend and participate.

WSSSPE3 also included two professional
event organizers/facilitators from Knowinnovation who helped the organizing committee members
plan the workshop agenda, and during the workshop, they actively engaged
participants with variously
tools, activities, and reminders. 

This report is based on collaborative notes taken during the workshop, which
were linked from the GitHub issues that represented the potential and actual
working
groups\footnote{\url{https://github.com/issues?q=label\%3A\%22WSSSPE3+activity\%22}}.
Overall, the report discusses the organization work done before the workshop
(\S\ref{sec:preworkshop}); the keynote (\S\ref{sec:keynote}); a series of
lightning talks (\S\ref{sec:lightning}). The report also gives summaries of
action plans proposed by the working groups (\S\ref{sec:WGs}), and some
conclusions (\S\ref{sec:conclusions}). Lists of the organizing committee
(Appendix~\ref{sec:orgcom}), the registered attendees
(Appendix~\ref{sec:attendees}), and the travel award recipients
(Appendix~\ref{sec:awardees}) are compiled. Finally, the report includes longer
descriptions of the activities that occurred in each of the working groups that
made substantial progress
(Appendices~\ref{sec:appendix_best_practices}--\ref{sec:appendix_user_community}).

\section{Calls for Participation} \label{sec:preworkshop}

WSSSPE3 was based on the work done in WSSSPE1 and WSSSPE2, but aimed at starting
a process to make progress in sustainable software, as the calls for
participation said:

\begin{quote} The WSSSPE1 workshop engaged the broad scientific community to
identify challenges and best practices in areas relevant to sustainable
scientific software. WSSSPE2 invited the community to propose and discuss
specific mechanisms to move towards an imagined future practice of software
development and usage in science and engineering. WSSSPE3 will organize
self-directed teams that will collaborate prior to and during the workshop to
create vision documents, proposals, papers, and action plans that will help the
scientific software community produce software that is more sustainable,
including developing sustainable career paths for community members. These teams
are intended to lead into working groups that will be active after the workshop,
if appropriate, working collaboratively to achieve their goals, and seeking
funding to do so if needed. \end{quote}

The first call for participation requested lightning talks, where each author
could make a brief statement about work that either had been done or was needed,
with the goal of contributing to the discussion of one or more working groups.
There were~24 lightning talks submitted; after a peer-review process, 16 
were accepted, as discussed further in Section~\ref{sec:lightning}.

The first call also discussed the potential action topics that came out of
WSSSPE2, and requested additional suggestions. The combination of existing and
new topics led to the following 18 potential topics that were advertised in the
subsequent calls for participation:
%
\begin{itemize}
\renewcommand{\labelenumi}{\textbf{\theenumi}.}
\setlength{\rightmargin}{1em}

\item Development and Community
\begin{itemize}
\item Writing a white paper/review paper about best practices in developing
sustainable software
\item Documenting successful models for funding specialist expertise in software
collaborations
\item Creating and curating catalogs for software tools that aid sustainability
(perhaps categorized by domain, programming languages, architectures, and/or
functions, e.g., for code testing, documentation)
\item Documenting case studies for academia/industry interaction
\item Determining effective strategies for refactoring/improving legacy
scientific software
\item Determining principles for engineering design for sustainable software
\item Create a set of guidance giving examples of specific metrics for the
success of scientific software in use, why they are chosen, what they are
useful to measure, and any challenges/pitfalls; then publish this as a white
paper
\end{itemize}

\item Training
\begin{itemize}
\item Writing a white paper on training for developing sustainable software, and
coordinating multiple ongoing training-oriented projects
\item Developing curriculum for software sustainability, and ideas about where
such curriculum would be presented, such as a summer training institute
\end{itemize}

\item Credit
\begin{itemize}
\item Hacking the credit and citation ecosystem (making it work, or work better,
for software)
\item Developing a taxonomy of contributorship/guidelines for including software
contributions in tenure review
\item Documenting case studies of receiving credit for software contributions
\item Developing a system of awards and recognitions to encourage sustainable
software
\end{itemize}

\item Publishing
\begin{itemize}
\item Developing a categorization of journals that publish software papers
(building on existing work), and case studies of alternative publishing
mechanisms that have been shown to improve software discoverability/reuse, e.g.,
popular blogs/websites
\item Determining what journals that publish software paper should provide to
their reviewers (e.g., guidelines, mechanisms, metadata standards)
\end{itemize}

\item Reproducibility and Testing
\begin{itemize}
\item Building a toolkit that could allow conference organizers to easily add a
reproducibility track
\item Documenting best practices for code testing and code review
\end{itemize}

\item Documentation
\begin{itemize}
\item Develop landing pages on the WSSSPE website (or elsewhere) that enable the
community to easily find up-to-date information on a WSSSPE topic (e.g.,
software credit, scientific software metrics, testing scientific software).
\end{itemize}

\end{itemize}

\section{Keynote}\label{sec:keynote}

WSSSPE3 began with a keynote speech delivered by Professor Matthew Turk from the
Department of Astronomy, University of Illinois, titled \emph{Why Sustain
Scientific Software?}. Turk is a prolific scientific software practitioner and
has extensive experiences working on large collaborative projects employing
modern computing tools \cite{2011ApJS..192....9T}. He also co-organizes and
champions WSSSPE events.

In his keynote address, Turk recapped the course of development of WSSSPE
workshops over the past few years, alongside his career development from a
postdoc to an academic. The first WSSSPE workshop was at the Supercomputing
conference (SC13) in 2013, but he observed that the notion of sustainable
scientific software drew in an audience beyond supercomputing. In the following
year, WSSSPE1.1 at SciPy had speakers talking about how software has been
sustained inside the scientific Python community. WSSSPE2 at SC14 had breakout
group discussions coming up with actionable items, and WSSSPE2.1 at SciPy 2015
was similar. Turk noted the different atmosphere of the surrounding large
conferences, despite similar WSSSPE participants.

WSSSPE3 left the traditional Supercomputing Conference environment this year,
and in Turk's words, this change spoke to the fact that scientific software
comes from many different types of inquiries, deployment, strategies for
maintenance, users, and ways of measuring the value of a piece of software. It
appeared to Turk that the supercomputing community generally adopts some
top-down approaches, whereas the SciPy community more often than not uses more
bottom-up systems. The essential messages perceived were also often bipolar: the
supercomputing community thinks that software is getting harder, with exascale
computing and optimization issues in mind; but the SciPy community thinks that
software is becoming better, with emerging tools such as Jupyter and
productivity packages for research workflows. Admitting such comparisons are
somewhat unfair generalizations, Turk reminded the audience that the different
approaches bring different types of ideas to the table, and he welcomed WSSSPE3
being conducted outside existing preconceptions.

Returning to the topic of his talk, Turk invited the audience to picture
scientific software as a flower on a landscape under the Sun, which may
represent a number of measurable factors such as number of citations; growth of
a community and number of contributors; amount of funding; prestigious prizes
awarded; stability of the community in terms of leadership transitions, serving
community needs, not breaking test suites, and performance on new architectures.
But all these metrics are strictly speaking \emph{proxies} for the values and
the impact scientific software bears. What we can measure does not give us
direct insight---it just gives us proxies of insight.
  
Turk then moved onto various different definitions of sustainability. His
favorite one was ``keeping up with bug reports,'' where even if no new features
were added, the software remains sustainable. Another definition of
sustainability Turk mentioned was ``adding of new features,'' or ``maintaining
the software for a long period of time'' such as the cases of \TeX\ or \LaTeX\
with community help. A notion Turk heard often at supercomputing conferences was
that sustainable software ``continues to work on new architectures.'' Yet
another metric was ``people continuing to be able to learn how to use and apply
the software.'' A funder Turk heard talked about sustainability as ``continuing
to get funded.'' Turk also recalled that Greg Wilson, among others, said in
WSSSPE1.1 that his view of sustainable software was software that ``continued to
give the same results over time.'' A last measure of sustainability Turk
presented was ``the ability to transition between different people developing
and using a piece of software.''

At WSSSPE1, several models were presented for ensuring sustainability. Turk
considered that a familiar one was a funded piece of software where an external
agency provided funds to a group who are not necessarily exclusively working on
and developing the software, keeps it going, and provides it to the scientific
community. The model of productized software, in which a piece of software has
grown to the point that research groups or people are willing to support it with
some amount of funding, for instance, a subscription to use cloud services that
deploy a piece of software, or purchase of a piece of software. A final model
Turk felt conflicted about is a volunteer model that is traditional
old-school---not modern-day open source---development.

Turk discussed whether productizing scientific software was synonymous with
being sustainable and self-sufficient. He thought it was not necessarily the
case and furthermore, it could lead to a divergence of interests between users
and developers.

Turk reminded the audience that the volunteer model means unpaid labor. On this
note, he recommended Ashe Dryden's blog post on the ethics of unpaid labor and
the open source software
community\footnote{\url{http://www.ashedryden.com/blog/the-ethics-of-unpaid-labor-and-the-oss-community}}.
Oftentimes, a person funded to work full time on a scientific project can spend
a small amount of time for working on a piece of software necessary for that
project. However, researchers' abilities to participate in that volunteer
community are not always the same and may not always be aligned with their
research projects. From Turk's experience, we cannot always rely on unpaid labor
and volunteer time to sustain a piece of software---this came down to the
notions of the top-down and the bottom-up approaches, i.e., the funded versus
the grassroots. However, Turk pointed out that bottom-up, volunteer-driven
projects can be just as large-scale as a top-down software development project.

Turk said that sustaining scientific software really meant to him conducting
scientific inquiries, often by some specific software, and sustaining the people
we care about, our careers, and the future of our fields. According to Turk, we
all have an invested stake in sustaining scientific software. Hence, having
``sustained'' projects can suffocate new projects, so we need to make sure we
don't cause novel ideas and packages to suffer at the hands of the status quo.

Turk talked about possible reasons why we want to sustain scientific software:
devotion to science and interests in pursuing the next stage of research; fun
and creative thrill in writing codes and papers; usefulness with measurable
impacts, for example, LINPACK and HDF groups providing data storage to
satellites, which goes beyond usefulness to necessity. Lastly, Turk presented
his wishlist of questions to be answered in the future:
\begin{itemize} 

\item How do we ship a product on time when dealing with a mix of funding models
and motivations especially when we rely on volunteers?

\item How do we know when it is time to end some software and move on? For
example, should we stop sustaining Python and switch to Julia and Javascript?
 
\item How can productized software balance its future versus its past, or the
emerging needs of the customers versus the existing needs of the development
community?

\item How can we help avoid burnout and retain the joy in the communities?

\item How can we reduce systemic bias, which goes back to Dryden's blog post
especially on how ethics of unpaid labor disproportionately affect
underrepresented communities?

\end{itemize}

\section{Lightning Talks} \label{sec:lightning}

The lightning talks were intended to give an opportunity for attendees to
quickly highlight an important issue or a potential solution.
\begin{enumerate}
\item \textbf{Benjamin Tovar and Douglas Thain: \textit{Freedom vs.\ Stability:
Facilitating Research Training While Supporting Scientific Research}} Benjamin
Tovar presented a case study of ``The Cooperative Computing Lab'' at the
University of Notre Dame, which is a small group of individuals whose main tasks
are collaborating with people that have large-scale computing problems,
operating various parallel computer systems, conducting computer science
research, and developing open source software. One of the main challenges they
face is finding a balance between flexibility/training and stability/quality.
Their current solution for ensuring the latter was to add a software engineer
(the presenter) to the existing team of faculty and students, who now also
serves as a ``spring'' between flexibility and stability.

\item \textbf{Birgit Penzenstadler, Colin Venters, Christoph Becker, Stefanie
Betz, Ruzanna Chitchyan, Let\'{i}cia Duboc, Steve Easterbrook, Guillermo
Rodriguez-Navas, and Norbert Seyff: \textit{Manifesting the Ghost of the Future:
Sustainability}} The concept of sustainability has become a topic of interest in
the field of computing, which is evidenced by the increase in the number of
events that focus on the topic. Nevertheless, it is not well understood yet.
Birgit Penzenstadler argued that we often define sustainability too narrowly.
Instead, sustainability at its heart is a systemic concept and must be viewed
from a range of different dimensions including environmental, economic,
individual, social, and technical. She introduced the Karlskrona Manifesto on
Software Design~\cite{Becker:2014}, which distills knowledge from a broad range
of related work on the topic of sustainability into a set of (mis-)perceptions
and principles. The manifesto does not proclaim that there is an easy,
one-size-fits-all solution around the corner, but rather points out that
sustainability is a ``wicked problem'' and is often misunderstood. Due to these
misperceptions, even though sustainability's importance is increasingly
recognized, many software systems are unsustainable. Even more alarming is that
most software systems' broader impacts on sustainability are unknown. To change
this, the Karlskrona Manifesto proposes nine principles and commitments. These
commitments are not dogmatic laws, but rather commitments to rethink, to move
beyond the silo mentality, and to analyze in more depth. As such, they do not
restrict, but rather open up a space for discussion.

\item \textbf{Abani Patra, Hossein Aghakhani, Nikolay Simakov, Matthew Jones,
and Tevfik Kosar: \textit{Integrating New Functionality Using Smart Interfaces
to Improve Productivity of Legacy Tools}} Abani Patra presented an example of
how the community using Titan2D, a geoflow simulation software, increased the
productivity of their tools by improving both code and data layout. The main
obstacles in this change were the non-existence of a common version control
system for the source code, coupled with multiple versions of the same code
base, the fixed format of input files, that many input values were set as
compilation flags, and that the internal data layout was not suitable for modern
technologies (e.g., vectorization, accelerators). The approaches of the Titan2D
developers included reinforcing the code structure using multiple layers of
Python and C++ interfaces, and a redesign of the data layout to be more suitable
for modern CPUs and accelerators.

\item \textbf{Abigail Cabunoc Mayes, Bill Mills, Arliss Collins, and Kaitlin
Thaney: \textit{Collaborative Software Development as Sustainable Software:
Lessons from Open Source}} Abigail Cabunoc Mayes combines two properties of open
source that, together, create a suitable habitat for sustainable software. The
first of the two properties, public, does not only mean public code. It also
includes public discussions, a public process of including contributions, and an
open license. The second property, participatory, stresses the importance of
reaching out to the community and helping potential new members by providing
better documentation and learning experiences, like code review and examples of
good first bug reports. Together, Abigail concluded, these two properties not
only lead to higher quality, reusability, and ease of understanding, but also
eventually, to sustainability.

\item \textbf{Louise Kellogg and Lorraine Hwang: \textit{Advancing Earth Science
through Best Practices in Open Source Software: Computational Infrastructure for
Geodynamics}} Lorraine Hwang presented experiences with the Computational
Infrastructure for Geodynamics (CIG), a community software with a worldwide user
base. Like others, their main goals include high usability, sustainability, and
reproducibility. As a means to achieve these goals, various communication
channels have been developed, such as mailing lists, wikis, workshops,
hackathons, tutorials, and webinars. In order to contribute to the
infrastructure, codes must adhere to specified minimum standards with the desire
that all codes are working toward target standards.
These include, e.g., the use of version control, certain coding styles, the
presence and nature of code tests and documentation, and certain user workflows.

\item \textbf{Lorraine Hwang, Joe Dumit, Alison Fish, Louise Kellogg, Mackenzie
Smith, and Laura Soito: \textit{Software Attribution for Geoscience Applications
in the Computational Infrastructure for Geodynamics}} In a second talk, Lorraine
Hwang mentioned a variety of ways to cite efforts within the framework, including
science papers, code papers, user manuals, and the CIG website. An analysis of
the resulting citations showed that $80\%$ of papers that use CIG codes mention
the code name, and about the same number includes a citation. Only about $20\%$
acknowledge CIG. Within the same sample of papers, about one fifth use an URL to
cite codes (including non-CIG codes), and only about one eighth specify the
version used. Compared to other codes, CIG seems to be much better cited. In
part, this is attributed to the fact that CIG requires that donated software
provide a citable paper specified in the User Manual. The project is working on
tools and methods to generate attribution information automatically.

\item \textbf{Mike Hildreth, Jarek Nabrzyski, Da Huo, Peter Ivie, Haiyan Meng,
Douglas Thain, and Charles Vardeman: \textit{Data And Software Preservation for
Open Science (DASPOS)}} DASPOS is an NSF-funded multi-disciplinary effort,
located at Notre Dame and Chicago, that links the high energy physics effort to
other disciplines such as biology, astrophysics, and digital curation. It includes
physicists, digital librarians, as well as computer scientists, and aims to
achieve some commonality across disciplines. Examples are meta-data descriptions
of archived data, computational descriptions, descriptions of how data was
processed, questions such as whether computation replication can be automated,
and what the impact of access policies on the preservation infrastructure is.
One of the products of this effort is a suite of tools that deals with this
preservation, and the questions was posed whether that software itself is
sustainable. Points that were brought up included the need for a user community
depending on a given software, and the need to provide added value for its
users. An important method to achieve this was to work with the user community
from the start, and to budget that way. For the specific example of the
preservation software, this means that besides adding value to the community, it
needs to be transparent to their workflows, i.e., not requiring additional
effort to preserve ``research objects.''

\item \textbf{James Hetherington, Jonathan Cooper, Robert Haines, Simon
Hettrick, James Spencer, Mark Stillwell, Mike Croucher, Christopher Woods, and
Susheel Varma: \textit{An update from UK Research Software Engineers}} James
Hetherington started by listing some of the problems research software faces,
which include poor standard of verification and low levels of reuse. For a long
time, technical solutions to such technical problems were focus of the eResearch
community, including research software distributions, grids, middleware and
workflows. Some limited adoption can be seen today in research communities, but
the main problems have not been solved to a sufficient level. Hetherington
hypothesized that instead of technical solutions, social innovation is needed: a
new role in the academic system focused on research software that combines the
best parts of a craftsperson and a scholar. However, social innovation in
centuries-old institutions is hard. Alternatives to such a new role would have
to include rewards for good research software, recognition of software as
academic output, and rejecting submissions based on irreproducible computational
results. Some advantages of research software engineering (RSE) groups include
the possibility of training in reproducible computational research, providing
collaborations for researchers who do not want to serve as programmers, and
creating synergies with research computing platforms. The success of such a
group would be measured by the output and quality of the research software.
Members could be part of Research Computing or faculty. They would not be
independent researchers, but would have to have a research background. An
attempt to form a community of RSE groups within the UK has been underway for
several years, including funding from the UK Research Council. An open question
is whether this approach can be adopted in other countries, including the~USA.

\item \textbf{Dan Gunter, Sarah Poon, and Lavanya Ramakrishnan: \textit{Bringing
the User into Building Sustainable Software for Science}} Dan Gunter's main
question was, ``What is needed to develop sustainable software?''. Beyond the
usual suspects of funding, proficient developers, good design, and software
engineering practices, Dan placed the users. He explained this using a
traditional software-development model starting from gathering requirements, and
reaching release through a design, development, and testing phase. The main
deficiency with this approach was pointed out to be the too-late interaction
with users. Instead, an alternative approach was proposed that, at first, skips
the development phase and repeatedly goes through requirement gathering, design
and user interaction/learning phases, and only eventually starts development
once an agreement is reached, leading to an increased user satisfaction, higher
adoption, and eventually to sustained software.

\item \textbf{Dan Gunter, Adam Arkin, Rick Stevens, Robert Cottingham, Sergei
Maslov, and the KBase Project: \textit{Challenges of a Sustainable Software
Platform for Predictive Biology: Lessons Learned on the KBase Project}} 
Dan Gunter presented experiences and lessons learned as part of KBase, an
open software and data platform for addressing the grand challenge of systems
biology: predicting and designing biological function. KBase is a unified system
that integrates data and analytical tools for comparative functional genomics of
microbes, plants, and their communities. However, it is also a collaborative
environment for sharing methods and results, and placing those results in the
context of knowledge in the field. Being a large, multi-institutional project,
one of the big challenges is to agree on standards to enable a single,
maintainable system. Working in isolation does not work (anymore) within this
field, and the community in the field also does not have standards for software
engineering. This is contrasted to computer science research, where software
engineering standards shorten design cycles, leading to more time for highly
rewarded activities like publishing, performance studies, graduating students,
or protecting ideas before publication. Instead of this more traditional
approach, KBase uses a variation on the ``Scrum'' methodology. After picking
projects and team members, four to five teams work on projects for about two
weeks before a one-week evaluation by the executive committee. Based on their
assessment new teams and projects might be chosen before iteratively restarting
the ``agile'' development cycle. This process is intentionally open and
documented.

\item \textbf{Yolanda Gil, Chris Duffy, Chris Mattmann, Erin Robinson, and Karan
Venayagamoorthy: \textit{The Geoscience Paper of the Future Initiative: Training
Scientists in Best Practices of Software Sharing}} Erin Robinson presented an
approach to overcome some hurdles in scientific publishing: disconnects among
experimental data, research software, and publications. A current effort in the
Geosciences is the ``Geoscience Paper of the Future,'' which includes four
elements. First, it forms a modern paper including text, data, and pointers to
supplementary materials. Second, it is reproducible, including data processing,
workflow, and visualization tools. Third, it is part of open science, which
includes being publicly available under open licenses, and providing meta-data.
Last but not least, it uses digital scholarship elements like persistent
identifiers for software, data, and authors, and it cites both data and
software. OntoSoft is a tool for helping with this effort, providing software stewardship
for the Geosciences. As part of this initiative, a special issue of a journal in
Geoscience areas is planned to include only Geoscience papers of the future,
with submissions open until the end of 2015. In addition, training sessions are
provided to geoscientists to learn best practices in software and data sharing,
provenance documentation, and scholarly publication.

\item \textbf{Neil Chue Hong: \textit{Building a Scientific Software
Accreditation Framework}} Neil Chue Hong presented a proposal to build a
scientific software accreditation framework. One of the aims of such a framework
would be to measure how ``good'' a given piece of software is, and to evaluate
how this can be effectively measured in the first place. This can be compared to
the effort of the standardized, easy to read, and understandable nutritional
labeling of food, which only contain a small set of categories. However, such a
framework for software would be more difficult due to different existing
community norms. The challenges such a framework faces include that many
measurements are subjective, that many metrics are too costly, and that
self-assessment needs to be encouraged. Possible categories would include
availability, usability, transferability, among others. Such a framework could
enable both improvement of specific software, as well as comparisons of similar
software. An accreditation by such a framework could then be part of software
management plans, ensuring that software is accessible and reusable throughout
the proposed project and beyond.

\item \textbf{Jeffrey Carver: \textit{On the Need for Software Engineering
Support for Sustainable Scientific Software}} Carver argued that for scientific
software to be truly sustainable, there is a need for developers to use
appropriate software engineering practices. His experience interacting with
scientific teams indicates that choosing and tailoring these practices is not a
trivial exercise. There is a general culture clash between software engineering
and science that hinders our ability to communicate and choose appropriate
methods. In addition, many experienced scientific software developers appear to
be unaware of software engineering practices that may be beneficial to them. The
most appropriate software engineering practices are those that are lightweight,
properly tailored, and focus on the key software development problems faced by
scientists. In order to increase the use of software engineering in science, we
need more documented success stories. These successes need to be socialized 
within the scientific community through workshops like the Software Engineering
for Science workshop
series\footnote{\href{http://www.SE4Science.org/workshops}{http://www.SE4Science.org/workshops}}
and the new Software Engineering track in \textit{Computing in Science and
Engineering} magazine.

\item \textbf{Matthias Bussonnier: \textit{User Data Collection in Open Source}}
This talk highlighted an attempt to solve the common problem for open source
development: it is difficult to collect information about how many people use
particular software, how often, which version, which parts of the software, or
on which operating system. Current solutions include surveys, but these have
high uncertainties. A different approach is based on automatic ``call-backs''
that collect these information at runtime and send it to a central place for
analysis. Problems with this approach include obtaining agreement from the user,
legal issues, increased maintenance (of servers), ethical questions, and also the
lack of a common infrastructure. Some of these problems are of a social nature
and have to be solved as such, but the last problem (a missing common
infrastructure) is attacked by the sempervirens project~\cite{sempervirens},
which is developing common APIs and a library implementation for common,
repeating tasks such as obtaining user consent. Results are uploaded not
directly to project servers, but to neutral third parties that only publish
aggregated statistics to projects.

\item \textbf{Alice Allen: \textit{We're giving away the store! (Merchandise not
included)}} Alice Allen described the Astrophysics Source Code Library
(ascl.net), an increasingly used way to obtain a unique ID for astrophysics
software that is indexed by indexing services and can be cited. ASCL offers
clones of existing infrastructure, provides server space and computing
resources, shares innovations, and permits moves elsewhere. Users provide a
domain name, then control and configure their site and use it as intended,
gather their codes as they wish, share innovations, and protect the provided
computing environment.

\item \textbf{Stan Ahalt, Bruce Berriman, Maxine Brown, Jeffrey Carver, Neil
Chue Hong, Allison Fish, Ray Idaszak, Greg Newman, Dhabaleswar Panda, Abani
Patra, Elbridge Gerry Puckett, Chris Roland, Douglas Thain, Selcuk Uluagac, and
Bo Zhang: \textit{Scientific Software Success: Developing Metrics While
Developing Community}} The effort behind this talk given by Ray Idaszak started
from a breakout group at an NSF SI2 workshop in 2015, and centers around
building a framework for creating metrics for scientific software. This
framework would improve both the metrics and the software it evaluates, and
could also serve as a tool for building a community around the idea. With
especially the last-mentioned idea (building a community) in mind, a software
``peer review group'' would be created, representing stakeholders who will
self-review software created by their respective communities, and will
concurrently develop metrics. The whole project should be community-governed,
without a single institution overseeing the activities or infrastructure, with
the hope to evolving community-generated and adopted standards. The generation
of metrics would be tied to the actual evaluation of software, creating an
incentive by improving the evaluated software itself during this process. The
framework code would provide infrastructure for the creation of metrics and
evaluation, and forums for generation of software success metrics. It would also
support code reviews of the evaluated software. An open question is whether it
is possible to fit the resulting metrics in a common template. So far, this is
still in a design phase, with a white paper at the CSESSP workshop 2015 and this
talk, but the WSSSPE workshops are seen as a forum for the community to assemble
and act, and is planned to be used also in the future to build this community
and framework.

\end{enumerate}

\section{Working Groups} \label{sec:WGs}


\subsection{White paper/journal paper about best practices in developing sustainable software}
\label{sec:best-practices}

Reviewing multiple past articles and talks at different meetings like WSSSPEx
\cite{WSSSPE1, WSSSPE2, 5069157, Blatt_WSSSPE, Ahern_WSSSPE} and analyzing and
promoting sustainable scientific software makes it clear that there are several
common and recurring ideas that underpin success in developing sustainable
software. However, outside of a small community, this knowledge is not widely
shared. This is especially true for the large community of scientists who
generate most of the software used by scientists but are not primarily software
developers. In this scenario, a clear and precise exposition of these best
practices collected from many sources and open collaboration among all in the
community in a single source (e.g., journal paper, tutorial) that can be widely
disseminated is necessary and likely to be very valuable.

\subsubsection{Fit with related activities}

The creation of such a ``best practices'' document will build upon the range of
activities and topics discussed at WSSSPE3 and associated prior meetings. We
will attempt to distill the emerging body of knowledge into this document. The
large number of articles from the NSF funded SI2 projects (SSE and SSI),
``lightning talks'', ``white papers,'' and reports from different workshops have
created a large if somewhat diffuse source for this report.

\subsubsection{Discussion}

Core questions that will need to be explored are in reproducibility, 
reliability,  usability, extensibility, knowledge management, as well as continuity
(transitions between people). Answers to these questions will guide us on how a software
tool becomes part of the core workflow of well identified users (stakeholders)
relating to tool success and hence sustainability.
Ideas  that may need to be explored include:
\begin{itemize}
\item Requirements engineering to create tools with immediate uptake;
\item When should software ``die''?
\item Catering to disruptive developments in environment (e.g., new hardware,
new methodology);
\item Dimensions of sustainability: economic, technical, environmental and
obsolescence. 
\end{itemize}

Sustainability requires community participation in code development and/or a
wide adoption of software. The larger the community base is using a piece of
software, the better are the funding possibilities and thus also the
sustainability options. Additionally developer commitment to an application is
essential and experience shows that software packages with an evangelist
imposing strong inspiration and discipline are more likely to achieve
sustainability. While a single person can push sustainability to a certain
level, open source software also needs sustained commitment from the developer
community. Such sustained commitments include diverse tasks and roles, which can
be fulfilled by diverse developers with different knowledge levels. Besides
developing software and appropriate software management with measures for
extensibility and scalability of the software, active (expertise) support for
users via a user forum with a quick turnaround is crucial. The barrier to entry
for the community as users as well as developers has to be as low as possible.

For additional information about the discussion, see
Appendix~\ref{sec:appendix_best_practices}.

\subsubsection{Plans}
The creation of a document on best practices needs a large and diverse community
involved. We have enlisted over ten contributors from the attendees at the
WSSSPE3 and those on the mailing list. The primary mechanism for developing this
document will be to examine and analyze the success of several well known
community scientific software and organizations supporting scientific software.
We will attempt then to abstract general principles and best practices. Some of
the tools identified for such analysis are the general purpose PETSc toolkit for
linear system solution, NWChem for computational chemistry and the CIG
(Computational Infrastructure for Geodynamics) organization dedicated to
supporting an ensemble of related tools for the geodynamics community. We also
established a timeline and a rough outline (see
Appendix~\ref{sec:appendix_best_practices}) for the report.

\medskip
\noindent{\bf Timeline:}
\begin{itemize}
\item 28 Dec: Introduction and scope finished
\item 06 Jan: Sections assigned
\item 31 Jan: Analyzing funding possibilities for survey
\item 31 Jan: First versions of section
\item 15 Feb: Distribution to WSSSPE community
\item 31 Mar: Final version of white paper
\item 30 Apr: Submission of peer-reviewed paper?
\end{itemize}

\subsubsection{Landing Page}
The landing page with instructions, timeline and the white paper is here: \url{https://drive.google.com/drive/folders/0B7KZv1TRi06fbnFkZjQ0ZEJKckk}.
Discussions can be also continued in \url{https://github.com/WSSSPE/meetings/issues/42}.

\subsection{Funding Research Programmer Expertise}
\label{RSE}



Research Software Engineers (RSEs)---those who contribute to science and
scholarship through software development---are an important part of the team
needed to deliver 21st century research. However, existing academic structures
and systems of funding do not effectively fund and sustain these skills. The
resulting high levels of turnover and inappropriate incentives are significant
contributing factors to low levels of reliability and readability observed in
scientific software. Moreover, the absence of skilled and experienced developers
retards progress in key projects, and at times causes important projects to fail
completely.

Effective development of software for advanced research requires that
researchers work closely with scientific software developers who understand the
research domain sufficiently to build meaningful software at a reasonable pace.
This requires a collaborative approach---where developers who are fully engaged
or invested in the research context are co-developing software with domain
academics.

\subsubsection{Fit with related activities}

The solution we envision entails creating an environment where software
developers are a stable part of a research team. Such an environment mitigates
the risk of losing a key developer at a critical moment in a projects lifetime,
and provides the benefits of building a store of institutional knowledge about
specific projects as well as about software development for today's research.
Our vision is to find a way to promote a university/research institute
environment where software developers are stable components of research project
teams.

One strategy to promote stability is implementing a mechanism for developers to
obtain academic credit for software development work. With such a mechanism in
place, traditional academic funding models and career tracks could properly
sustain individuals for whom software development is their primary contribution
to research. A contributing factor to the problem with the current academic
reward system is the devastating effect on an academic publication record
resulting from time in industry; such postings often develop exactly the skills
that research software engineers need, yet returns to university positions
following an industry role are penalized by the current structures. Retention of
senior developers is hard, because these people are high in demand by the
economy. However, people who have a PhD in science and enter industry, may
desire to return for diverse reasons, and should be welcomed back.

While developing new mechanisms in the current academic reward system is a
worthy aspirational goal, such a dramatic change in this structure does not seem
likely in a time scale relevant to this working group. Accordingly, our working
party sought alternative solutions that may be achievable within the context of
existing academic structures. The group felt that developing dedicated research
software engineering roles within the university, and finding stable funding for
those individuals is the most promising mechanism for creating a stable software
development staff.

Measures of impact and success for research programming groups, as well as for
individual research software engineers, will be required in order to make the
case to the university for continued funding. Research software engineers will
not be measured by publications, we hope, but by other metrics. Middle-author
publications are common for RSEs. Most RSEs welcome co-authorship on papers
where the PIs think that the contribution deserves it.

\subsubsection{Discussion}

It is hard for an individual PI in a university or college to support dedicated
research software engineering resources, as the need and funding for these
activities are intermittent within a research cycle. To sustain this capacity,
therefore, it is necessary to aggregate this work across multiple research
groups.

One solution is to fund dedicated software engineering roles for major research
software projects at national laboratories or other non-educational
institutions. This solution is in place and working well for many well-used
scientific codebases. However, this strategy has limited application, as much of
the body of software is created and maintained in research universities.
Therefore, we argue that research institutions should develop hybrid
academic-technical tracks for this capacity, where employees in this track work
with more than one PI, rather than the traditional RA role within a single
group. This could be coordinated centrally, as a core facility, perhaps within
research computing organizations which have traditionally supported university
cyberinfrastructure, library organizations, or research offices. Alternatively,
these groups could be organizationally closer to research groups, sitting within
academic departments. The most effective model will vary from institution to
institution, but the mandate and ways of working should be similar.

Having convinced ourselves that this would be a positive innovation, we were
then faced with the specific question of how to fund the initiation of this
activity. A self-sustaining research software group will support itself through
collaborations with PIs in the normal grant process, with PIs choosing to fund
some amount of research software engineering effort through grants in the usual
way. However, to bootstrap such a function to a level where it has sufficient
reputation and client base to be self-sustaining will generally require seed
investment.

This might come from universities themselves (this was the model that led to the
creation of the group in University College London), but more likely, seed
funding needs to come from research councils (as with the Research Software
Engineering Fellowship provided by the UK Engineering and Physical Sciences
Research Council). We therefore recommend that funding organizations consider
how they might provide such seed funding.

Success, appropriately measured, will help make the case to such funding bodies
for further investment. One might expect that metrics such as improved
productivity, software adoption rates, and grant success rates would be
sufficient arguments in favor of such a model. However, useful measurement of
code cleanliness, and the resulting productivity gains, is an unsolved problem
in empirical software engineering. To measure ``what did not go wrong'' because
of an intervention is particularly hard.

We finally noted that the institutional case for such groups is made easier by
having successful examples to point to. In the UK, a collective effort to
identify the research software engineering community, with individuals clearly
stating ``I am a research software engineer,'' has been important to the
campaign. It will be useful to the global effort to similarly identify emerging
research software organizations, and also, importantly, to identify
longer-running research software groups, which have in some cases had a long
running \emph{sui-generis} existence, but which now can be identified as part of
a wider solution. There remains the problem of how to ``sell'' the value of this
investment to investigators within a university. This is an issue best addressed
by the individual organizations that embark on the plan.

For more details on the discussion, see
Appendix~\ref{sec:appendix_funding_spec_expert}.

\subsubsection{Plans}

The first step in moving this strategy forward is to gather a list of groups
that self-identify as research software engineering groups, and to reach out to
other organizations to see if there may be a widespread community of RSEs who do
not identify themselves as such at this time. We will collect information about
the organizational models under which these groups function, and how they are
funded. For example, how many research universities currently fund people in the
RSE track, whether they bear the RSE moniker or not. Are these developers paid
by the university or through a program supported by research grants/individual
PIs? How did they bootstrap the developer track to get this started? How
successful is the university in getting investigators to pay for fractional
RSEs? We will author a report describing our findings, should funding be
available to conduct the investigation.

\subsubsection{Landing Page}

A list of known UK research software engineering groups is available at  
\url{http://www.rse.ac.uk/groups}, and a list for the rest of the world
is at \url{http://www.rse.ac.uk/international}.
To add another group to the list, please make a pull request as requested
on either of these pages.


\subsection{Transition Pathways to Sustainable Software: Industry \& Academic Collaboration}
\label{sec:industry_interaction}


Most scientific software is produced as a part of grant-funded research projects
typically sponsored by federal governments. If we are interested in the
sustainability of scientific software, then we need to understand what exactly
happens when that sponsorship ends. More than likely, the project and its
resulting software will need to undergo some kind of transition in funding and
consequently governance.

At WSSSPE3 our group was interested in better understanding successful pathways
for scientific software to ``transition'' from grant-funded research projects to
industry sponsorship. (This may be an initially awkward phrase---some software
projects will begin their life being sponsored by industry, or result in
collaboration between industry and academia. In such cases, there is still a
need to understand how IP and how maintenance of the software is sustained over
time.)

\subsubsection{Fit with related activities}

Most previous research and discussion of industry and academic collaboration,
sharing, and funding of research software has focused on the impact of such
arrangements. Examples of these types of reports are:

\begin{itemize}
\item REF Impact Case Studies: \url{http://impact.ref.ac.uk/CaseStudies/}
\item Background of projects funded in the UK: \url{http://gtr.rcuk.ac.uk/}
\item Dowling Review from the UK: addresses complexity of work between these two
communities: \url{http://www.raeng.org.uk/policy/dowling-review}
\item Pathway to Impact -- UK report: two pages of grant proposals are asked to
forecast what impact they might have (including environmental, academic, economic).
\end{itemize}

\subsubsection{Discussion}

Although sustainability transitions are often studied under the broad umbrella
of ``technology transfer,'' we believe there are likely to be a number of
different ways in which a pathway from initial production to long-term
maintenance and secure funding is achieved. In short, industry sponsorship
and/or direct participation is an important aspect of sustaining scientific
software, but our current understanding of these transitions focuses narrowly on
commercial successes or failures of those collaborations.

In looking at existing literature that addresses industry transitions, many
reports (such as those listed above) focus on benefits that accrue to the
private sector, or to a government that originally sponsored the research
project. This literature does not address the impact that these transitions have
on the accessibility or usability of the software, or the impact that these
transitions have on the career of the researchers involved.

For more detail on the group's discussion, see
Appendix~\ref{sec:appendix_industry_interaction}.

\subsubsection{Plans}

Plans for carrying forward are currently unclear---this project would require
sustained attention and effort from our team, and at least some amount of
funding in order for us to be involved for extended periods of time.

The broad goals that we would like to accomplish are: 

\begin{enumerate}

\item To complete a set of case studies which look at successful and
unsuccessful transitions between academic researchers and industry

\item To create a generalizable framework, which might allow for a broader study
of different transition pathways (other than between academia and industry)

\end{enumerate}

The main plan for the group going forward is the creation of a white paper on
the topic of sustainability transitions.

\subsubsection{Landing Page}

Transitions Pathways discussions can be posted at
\url{https://github.com/WSSSPE/meetings/issues/46} or an email be sent to Nic
Weber\footnote{email: \href{mailto:nmweber@uw.edu}{nmweber@uw.edu}} to find out
more about the group's efforts and how to participate.

\subsection{Legacy Software} 

This group met only briefly, for one period on the first day. They discussed
that it is difficult to define legacy code because there is so much stigma
associated with the term. At some point there will be more difficulty and
resources wasted trying to keep legacy software supported, but it will
eventually be too expensive compared to how much it would be to just rebuild the
software or kill it. Most of the group members were not able to attend on the
second day, and those who were able to attend joined other groups.

%
%
%
%

\subsection{Principles for Software Engineering Design for Sustainable Software} 


Principles for software engineering form the basis of methods, techniques,
methodologies and tools~\cite{Vliet:2008}. However, there is often a mismatch
between software engineering theory and practice particularly in the fields of
computational science and engineering, which can lead to the development of
unsustainable software~\cite{Merali:2010,hettrick:2014}. Understanding and
applying software engineering principles is essential in order to create and
maintain sustainable software~\cite{Becker:2016}.

\subsubsection{Fit with related activities}

The group discussion focused on identifying existing principles of software
engineering design that could be adopted by the computational science and
engineering communities.

\subsubsection{Discussion}

Software engineering principles form the foundation of methods, techniques,
methodologies, and tools.  Consisting of members from different
backgrounds, including quantum chemistry, epidemiology, computer science,
software engineering, and microscopy, this group discussed the principles of software
engineering design for sustainable software (starting with principles from the
Karlskrona Manifesto on Sustainability Design~\cite{Becker:2015},
Tate~\cite{tate:2005}, and the Software Engineering Body of Knowledge
(SWEBOK)~\cite{swebokv3}) and their application in various domains including
quantum chemistry and epidemiology. The group examined the principles and took a
retrospective analysis of what the developers did in practice against how the
principles could have made a difference, and asked, what do the principles mean
for computational scientific and engineering software, and how do the principles
relate to non-functional requirements? It appeared that the sustainable software
engineering principles should be mapped to two core quality attributes that
underpin technically sustainable software: \emph{extensibility}, the software's ability
to be extended and the effort level required to implement the extension; and
\emph{maintainability}: the effort required to locate and fix an error in operational
software.

For more information about the discussion, see Appendix~\ref{sec:appendix_eng_design}.

\subsubsection{Plans}

The next steps in this endeavor are to (1) Systematically analyze a number of
example systems from different scientific domains with regards to the identified
principles, to (2) Identify the commonalities and gaps in applying those
principles to different scientific systems, and to (3) Propose a set of
guidelines on the principles and examine how they exemplarily apply to scientific software
systems. Preliminary work will be carried out through undergraduate or
post-graduate student projects.

\subsubsection{Landing Page}

In the absence of a landing page, the Principles for Software Engineering Design
for Sustainable Software working group requests an email be sent to Birgit
Penzenstadler\footnote{email:
\href{mailto:birgit.penzenstadler@csulb.edu}{birgit.penzenstadler@csulb.edu}}
and Colin C.\
Venters\footnote{email:\href{mailto:c.venters@hud.ac.uk}{c.venters@hud.ac.uk}}
to find out more about the group's efforts and how to participate.

\subsection{Useful Metrics for Scientific Software}
\label{sec:software-metrics}


Metrics for scientific software are important for many purposes, including tenure and promotion,
scientific impact, discovery, reducing duplication, serving as a basis for
potential industrial interest in adopting software, prioritizing development and
support towards strategic objectives, and making a case for new or continued
funding. However, there is no commonly-used standard for collecting or
presenting metrics, nor is it known if there is a common set of metrics for
scientific software. It is imperative that scientific software stakeholders
understand that it is useful to collect metrics.

\subsubsection{Fit with related activities}

The group discussion focused on identifying existing frameworks and activities
for scientific software metrics. The group identified the following related
activities:

\begin{itemize}

\item Computational Infrastructure for Geodynamics: Software Development Best
Practices\footnote{\url{https://geodynamics.org/cig/dev/best-practices/}}

\item WSSSPE3 Breakout Session: How can we measure the impact of a piece of code on
research, and its value to the
community?\footnote{\url{https://docs.google.com/document/d/1cgUDH3RxrfsLotWhKKOrXUnaYFhrtjcV1TDRkFtwQKI/edit}}

\item 2015 NSF SI2 PI Workshop Breakout Session on Framing Success
Metrics\footnote{\url{https://docs.google.com/document/d/10yj7MYEjvrg__t522XR41ogASYMp647-l-BpFTsqEV4/}}

\item 2015 NSF SI2 PI Workshop Breakout Session on Software
Metrics\footnote{\url{https://docs.google.com/document/d/1uDim5bw8rBuubmtaUrz5Eh35NxzDgivmmdXhVzDs3tc/edit}}

\item NSF Workshop on Software and Data Citation Breakout Group on Useful
Metrics\footnote{\url{https://docs.google.com/presentation/d/1PPLVL6uoOmisqnHTlwhsVKJBTFFK1IVzvr8FdEEIvAE/}}

\item U.K. Software Sustainability Institute Software Evaluation
Guide\footnote{\url{http://www.software.ac.uk/software-evaluation-guide}}

\item U.K. Software Sustainability Institute Blog post: The five stars of
research
software\footnote{\url{http://www.software.ac.uk/blog/2013-04-09-five-stars-research-software}}

\item Minimal information for reusable scientific
software\footnote{\url{http://figshare.com/articles/Minimal_information_for_reusable_scientific_software/1112528}}

\item EPSRC-funded Equipment Data Search
Site\footnote{\url{http://equipment.data.ac.uk/}}

\item Canarie Research Software: Software to accelerate
discovery\footnote{\url{http://www.canarie.ca/software/}}

\item Canarie Research Software: Research Software Platform
Registry\footnote{\url{https://science.canarie.ca/researchmiddleware/platforms/list/main.html}}

\item BlackDuck Open HUB\footnote{\url{https://www.openhub.net/}}
\item Innovation Policy
Platform\footnote{\url{https://www.innovationpolicyplatform.org/frontpage}}

\end{itemize}

\subsubsection{Discussion}

The group discussion began by agreeing on the common purpose of creating a set
of guidance giving examples of specific metrics for the success of scientific
software in use, why they were chosen, what they are useful to measure, and any
challenges and pitfalls; then publish this as a white paper. The group discussed
many questions related to useful metrics for scientific software including
addressing if there is a common set of metrics that can be filtered in some way,
can metrics be fit into a common template, which metrics would be the most
useful for each stakeholder, which metrics are the most helpful and how would we
assess this, how are metrics monitored, and many more. A more complete bulleted
list of these questions can be found in Appendix H. Next, a roadmap for how to
proceed was discussed including creating a set of milestones and tasks. The idea
was put forth for the group to interact with the organizing committee of the
2016 NSF Software Infrastructure for Sustained Innovation (SI2) PI workshop in
order to send a software metrics survey to all SI2 and related awardees as a
targeted and relevant set of stakeholders. The five solicitations for software
elements released under the NSF SI2 program all included metrics as a required
component with submitters requested to include {\it ``a list of tangible
metrics, with end user involvement, to be used to measure the success of the
software element developed, \dots''}. These metrics are then reported as part of
annual reports to NSF by the projects. Although neither the proposal text
describing the metrics nor the reported metric results are publicly available,
there is reason to believe that the community will be willing to provide this
information through a survey mechanism. This survey would be created by one of
the student group members. Similarly, it was suggested that a software metrics
survey be sent to the UK SFTF (Software For The Future, led by the Engineering
and Physical Sciences Research Council) and TRDF (Tools and Resources
Development Fund, led by the Biotechnology and Biological Sciences Research
Council) software projects to ask them what metrics would be useful to report.
The remainder of the discussion focused mainly on the creation of a white paper
on this topic. This resulted in a paper outline and writing assignments with the
goal of publishing in venues including WSSSPE4, IEEE CiSE (Institute of
Electrical and Electronic Engineers Computing in Science and Engineering
magazine), or JORS (Journal of Open Research Software). More information about
the group discussion is available in Appendix~\ref{sec:appendix_metrics}.

\subsubsection{Plans}

The main plan for the group going forward is the creation of a white paper on
the topic of useful metrics for scientific software. The authoring of this white
paper would happen in parallel with the creation of a survey by the group with
the survey results to be incorporated in the white paper. The timeline for
completion of the white paper is approximately one year targeting venues
discussed in the previous section.

\subsubsection{Landing Page}

In lieu of a landing page, the Useful Metrics for Scientific Software working
group requests an email be sent to Gabrielle Allen\footnote{email: 
\href{mailto:gdallen@illinois.edu}{gdallen@illinois.edu}} to find out more
about the group's efforts and how to participate.

\subsection{Training}


This group explored a rapidly growing array of training that is seen to
contribute to sustainable software. The offerings are diverse, providing
training that is more or less directly relevant to sustainable software. While
research institutions support professional development for research staff, the
skills taught which might impact on sustainable software are limited at best,
often lacking a clear and coherent development pathway. Bringing together those
involved in leading relevant initiatives on a regular basis could helpfully
coordinate this growing array of training opportunities.

\subsubsection{Fit with related activities} 

Three existing venues for discussion of related events are
identified:
\begin{itemize}

\item Working towards Sustainable Software for Science: Practice and
Experiences (WSSSPE) workshops~\cite{WSSSPE}

\item International Workshop on Software Engineering for High
Performance Computing in Computational Science and
Engineering (SEHPCCSE)~\cite{SEHPCCSE}

\item Workshop on Software Engineering for Sustainable Systems~\cite{se4susy}

\end{itemize}

\subsubsection{Discussion}

Some next steps were identified to quickly test whether there is interest in
establishing a community committed to increasing the degree of coordination
across training projects. See Appendix~\ref{sec:appendix_training} for more
details about the discussion.

\subsubsection{Plans}

The main plan for the group is to convene a discussion to explore bringing
together regular meetings of those involved in leading relevant training
projects.

\subsubsection{Landing Page}

The Training working group requests an email be sent to Nick Jones\footnote{email: 
\href{mailto:nick.jones@nesi.org.nz}{nick.jones@nesi.org.nz}} to find out more
about the group's efforts and how to participate.

\subsection{Software Credit Working Group}
\label{sec:software-credit}


Modern scientific and engineering research often relies considerably on software, 
but currently no standard mechanism exists for citing software or receiving 
credit for developing software akin to receiving credit via citations for 
writing papers. Ensuring that developers of such scientific software receive 
credit for their efforts will encourage additional creation and maintenance. 
Standardizing software citations offers one route to establishing such a
citation and credit mechanism. Software is currently eligible for DOI
assignment, but DOI metadata fields are not well tuned for software compared to
publications. Some software providers apply for DOIs but it is still not widely
adopted. Also, there is no mechanism to cite software dependencies within
software in the same way papers cite supporting prior work.

\subsubsection{Fit with related activities}

Publishing Software Working Group (\S\ref{sec:publishing-software}): publishing
a software paper offers one existing mechanism for receiving credit, and further
developing new publishing concepts for software will strengthen our activities.

A number of groups external to WSSSPE (although with some overlapping members)
are also focused on aspects of software credit, including the FORCE11 Software
Citation Working Group (see plans for coordination below). In addition, a
Software Credit workshop\footnote{London Software Credit workshop:
\url{http://www.software.ac.uk/software-credit}} convened in London on October
19, following the conclusion of WSSSPE3. See
Appendix~\ref{sec:appendix_SW_credit} for more detailed discussion of related
activities.

\subsubsection{Discussion}

The group discussed a number of topics related to software credit, including a
contributorship taxonomy, software citation metadata, standards for citing
software in publications, and increasing the value of software in academic
promotion and tenure reviews. Although initial discussions both prior to and
during WSSSPE3 focused on contribution taxonomy and dividing credit, discussing
as an example the Entertainment Identifier Registry~\cite{EIDR} used in the
entertainment industry, the group decided to prioritize software citation. This
decision was motivated by the idea that standardizing citations for software
would introduce some initial credit for developers, and later the quantification
of credit could be refined based on concepts such as transitive
credit~\cite{wssspe2_katz,Katz:2014_tc}.

The majority of the remaining discussion focused on standardizing (1) the
metadata necessary for software to be cited and (2) the mechanism for citing
software in publications. Moreover, discussions also oriented around the
indexing of software citations necessary for establishing a software citation
network either integrated with the existing paper citation ecosystem or
complementary to it. See Appendix~\ref{sec:appendix_SW_credit} for a more
detailed summary of the working group's discussion on these topics.

\subsubsection{Plans}

The group already merged with the FORCE11 Software Citation Working Group
(SCWG), and their efforts will focus (over the next six to nine months) on
developing a document describing principles for software citation. Following the
publication of that document, the group will focus on outreach to key
groups (e.g., journals, publishers, indexers, professional societies).
Longer-term plans include working with indexers to ensure that software
citations are indexed and pursuing an open\slash community indexer; these
activities may be organized by future FORCE11 working groups.

\subsubsection{Landing Page}

Since near-term efforts will be shifting to the FORCE11-SCWG, we direct
interested readers to that group's existing landing page\footnote{FORCE11-SCWG
landing page,
\url{https://www.force11.org/group/software-citation-working-group}} and GitHub
repository\footnote{FORCE11-SCWG GitHub page,
\url{https://github.com/force11/force11-scwg}}.

\subsection{Publishing Software Working Group Discussion} \label{sec:publishing-software}

This working group explored the value of executable papers (papers whose content
includes the code needed to produce their own results), and other forms of
publishing which include dynamic electronic content.
%
%
Transitioning to this type of publication offers possibilities of addressing, or
partially addressing, sustainability concerns such as reproducibility, software
credit, and best practices.

\subsubsection{Fit with related activities}

\begin{itemize}
\item \textbf{Reproducibility}: Part of the purpose of these executable paper venues
is to (at least partially) address the reproducibility issue by making papers
recompute their own results.

\item\textbf{Software Credit (\S\ref{sec:software-credit})}: Since these forms of
publishing must make their sources explicit in order to execute, they should be
easier to trace even if appropriately worded credit for software is not
provided. In addition, they
make it possible to provide or define
additional metadata to make the tracing of credit clearer. Finally, attributions
could be added to citations to identify whether a paper extends a result,
verifies it, contradicts it, etc.

\item \textbf{Best Practices (\S\ref{sec:best-practices})}: Because an executable
paper showcases the code, and the code itself is subject to the review process,
authors are more likely to pay attention to coding practices. In addition,
because the paper must explain what the code does, better documentation is
more likely to be achieved.
\end{itemize}

\subsubsection{Discussion}

The group felt that the best way to encourage the use of these new publishing
concepts would be to create and curate a list of publishing venues that
support them. The Software Sustainability Institute agreed to host this list.

See Appendix~\ref{sec:appendix_publishing_SW} for more details about the
discussion.

\subsubsection{Plans}

The plan is to create and curate a web page describing executable papers, their value, and
a list of what publishers support them. We expect the page to be available in
early January of 2016 on the Software Sustainability Institute's website.

\subsubsection{Landing Page}

The aforementioned page will be published on the Software Sustainability
Institute website: \url{http://www.software.ac.uk}.

\subsection{Building Sustainable User Communities for Scientific Software}


User communities are the lifeblood of sustainable scientific software. The user
community includes the developers, both internal and external, of the software;
direct users of the software; other software projects that depend on the
software; and any other groups that create or consume data that is specific to
the software. Together these groups provide both the reason for sustaining the
software and, collectively, the requirements that drive its continued evolution
and improvement.

\subsubsection{Fit with related activities}

There are a number of activities already in progress that are targeted at improving
the user community for open-source software, including Mozilla Science's ``Working Open Project
Guide''~\cite{working-open-wssspe3} and
``UK Collaborative Computational Projects'' (CCP: http://www.ccp.ac.uk), or 
 books such as ``Art of Community'' by Jono Bacon~\cite{art-of-community}. 

\subsubsection{Discussion}

Discussion revolved around a few questions: what are the benefits of having a
``community'' for software sustainability; what practices and circumstances may lead
to having and maintaining a community; how can funding help or hinder this
process; and perhaps most importantly, how can best practices be described and
distilled into a document that can help new projects.

Everyone agreed on a few points: software must not only offer value, but there
must be some support for users; and funding can help pay for that support, in
addition to the usual funding for software development. Openness is generally 
a virtue. An evangelist, either in the form of a single person or some
domain-specific group of users, is often the key factor.

Additional details on the group's discussion can be found in
Appendix~\ref{sec:appendix_user_community}.

\subsubsection{Plans}

The most important next steps is a ``Best Practice'' document, which would
describe what successful projects with engaged communities look like, how to
replicate this type of project, and look at the end of life of a community project.
Another next step would be better training to increase recognition of need for science software
projects to focus on building and supporting their user communities.

\subsubsection{Landing Page}

This group does not have a landing page yet. Please send requests to join and
contribute by writing to both Dan Gunter\footnote{email:
\href{mailto:dkgunter@lbl.gov}{dkgunter@lbl.gov}} and Ethan
Davis\footnote{email: \href{mailto:edavis@ucar.edu}{edavis@ucar.edu}}.

\section{Conclusions} \label{sec:conclusions}

In WSSSPE3, we attempted to take what we learned from WSSSPE1 and WSSSPE2 in how
we can collaboratively build a workshop agenda and turn that into an ongoing
community activity. The success or failure of these efforts will
only become apparent over time.

The workshop had two components, presentations and working groups. The
presentations, in the first half day of the workshop, included an inspirational
keynote and a set of lightning talks. We used lightning talks for two reasons:
first, the need of some participants to have a slot on the agenda to justify
their attendance; and second, as a way to get new ideas across to all the
attendees. We broke with the tradition of requiring the lightning talk
submitters to self-publish their papers, and instead used a common peer-review
platform\footnote{\url{http://easychair.org}}, choosing to publish their slides
on the workshop website instead.

The working groups met for a small part of the first half day and all of the
second day, with the exception of some short periods for the groups to report
back to the collected workshop attendees. Each group determined a set of
activities that the members could do to advance sustainable software in a
particular area.

The results of these group sessions made it clear that there are many
interlinked challenges in sustainable software, and that while these challenges
can be addressed, doing so is difficult because they generally are not the
full-time job of any of the attendees. As was the case in WSSSPE2 as well, the
participants were willing to dedicate their time to the groups while they were
at the meeting, but afterwards, they went back to their (paid) jobs.

We need to determine how to tie the WSSSPE breakout activities to people's jobs,
so that they feel that continuing them is a higher priority than it is now,
perhaps through funding the participants, or through funding coordinators for
each activity, or perhaps by getting the workshop participants to agree to a
specific schedule of activities during the workshop as we have tried to do in
WSSSPE3. It remains to be seen, however, if the participants will meet the
schedules they set.

The overall challenge left to the sustainable software community is perhaps one
of organization: how to combine the small partial efforts of a large number of
people to impact a much larger number of people: those who develop and use
scientific software. While WSSSPE might help focus the actions of the groups,
something more is needed to incentivize the wider community, which is a
generalization of the sustainable software problem itself.

\section*{Acknowledgments} \label{sec:acks}

Work by Katz was supported by the National Science Foundation while working at
the Foundation. Any opinion, finding, and conclusions or recommendations
expressed in this material are those of the author(s) and do not necessarily
reflect the views of the National Science Foundation.
Choi's work was supported in part by the National Science Foundation research
grant DMS-1522687 and a WSSSPE3 travel award. She thanks the encouragement and
discussion with Fred Hickernell.
Hetherington was funded by the Software Sustainability Institute, RCUK grants
EP/H043160/1 and EP/N006410/1.
Work by Gunter was supported by the Office of Science, Office of Biological and
Environmental Research, of the U.S.\ Department of Energy (DOE) under Award
Numbers DE-AC02-05CH11231, DE-AC02-06CH11357, DE-AC05-00OR22725, and
DE-AC02-98CH10886, as part of the DOE Systems Biology Knowledgebase and by the
Office of Science, Office of Advanced Scientific Computing Research (ASCR) of
the U.S.\ Department of Energy under Contract Number DE-AC02-05CH11231 as part
of the Template Interfaces for Agile Parallel Data-Intensive Science (TIGRES)
project.
%

\newpage
\appendix
\section{Organizing Committee}  \label{sec:orgcom}

The following is the list of organizers of WSSSPE3.

{\scriptsize
\begin{longtable}{lll}
Daniel S. Katz &  University of Chicago \& Argonne National Laboratory, USA\\
Gabrielle Allen &  University of Illinois Urbana-Champaign, USA\\
Neil Chue Hong &  Software Sustainability Institute.  University of Edinburgh, UK\\
Sou-Cheng (Terrya) Choi &  NORC at the University of Chicago \& Illinois Institute of Technology, USA\\
Sandra Gesing &  University of Notre Dame,  USA\\
Lorraine Hwang &   University of California, Davis, USA\\
Manish Parashar &  Rutgers University, USA\\
Erin Robinson &  Foundation for Earth Science, USA (local organizer)\\
Matthew Turk &  University of Illinois Urbana-Champaign, USA\\
Colin C. Venters &  University of Huddersfield, UK

\end{longtable}
}

\section{Attendees}  \label{sec:attendees}
The following is a list of participants registered for the WSSSPE3 workshop.

{\scriptsize
\begin{longtable}{lll}
Alice Allen & Astrophysics Source Code Library\\
Gabrielle Allen & NCSA\\
Janine Aquino & UCAR/NCAR Earth Observing Laboratory\\
Steven Brandt & Louisiana State University\\
Jed Brown & CU Boulder\\
Matthias Bussonnier & UC Berkeley\\
Jeffrey Carver & University of Alabama\\
Emily Chen & NCSA\\
Sou-Cheng Choi &  NORC at the University of Chicago \&  Illinois Institute of Technology\\
Nancy Collins & NCAR\\
Ethan Davis & UCAR Unidata\\
Davide DelVento & NCAR/CISL\\
Yuhan Ding & Illinois Institute of Technology\\
Tim Dunne & KnowInnovation \\
Ward Fisher & UCAR/Unidata\\
Sandra Gesing & University of Notre Dame\\
Josh Greenberg & Sloan Foundation\\
Dan Gunter & LBNL\\
Ted Habermann & The HDF Group\\
James Hetherington & University College London\\
Neil Chue Hong & Software Sustainability Institute\\
Elisabeth Huffer & Lingua Logica/NASA \\
Lorraine Hwang & UC Davis - CIG\\
Raymond Idaszak & RENCI; University of North Carolina at Chapel Hill\\
Elizabeth Jessup & University of Colorado Boulder\\
Nick Jones & New Zealand eScience Infrastructure (NeSI)\\
Daniel Katz & U Chicago \& Argonne\\
Iain Larmour & EPSRC (UK)\\
Frank L\"offler & Louisiana State University\\
Suresh Marru & Indiana University\\
Ryan May & UCAR/Unidata\\
Abigail Cabunoc Mayes & Mozilla Foundation\\
Jeff McWhirter & Geode Systems\\
Constantinos Michailidis & Knowinnovation\\
Don Middleton & NCAR\\
Mark Miller & SDSC\\
Pate Motter & University of Colorado\\
Jaroslaw Nabrzyski & University of Notre Dame\\
Patrick Nichols & National Center for Atmospheric Research\\
Kyle Niemeyer & Oregon State University\\
Laura Owen & NCSA\\
Abani Patra & Univ at Buffalo\\
Grace Peng & National Center for Atmospheric Research\\
Birgit Penzenstadler & California State University Long Beach\\
Lindsay Powers & The HDF Group\\
Bernie Randles & UCLA\\
Erin Robinson & Foundation for Earth Science\\
Daniel Sellars & CANARIE Inc\\
Nikolay Simakov & SUNY University at Buffalo\\
Ian Taylor & Cardiff University\\
Ilian Todorov & Science \& Technology Facilities Council, UK\\
Benjamin Tovar & University of Notre Dame\\
Gregory Tucker & University of Colorado at Boulder\\
Matthew Turk & NCSA\\
Colin Venters & University of Huddersfield\\
Alexander Vyushkov & University of Notre Dame\\
Fraser Watson & National Solar Observatory\\
Nic Weber & University of Washington\\
Daniel Ziskin & NCAR - ACOM\\

\end{longtable}
}

\section{Travel Award Recipients}  \label{sec:awardees}
The following is the list of travel award recipients for the WSSSPE3 workshop.

{\scriptsize
\begin{longtable}{lll}
Alice Allen & Astrophysics Source Code Library\\
Steven Brandt &  Louisiana State University\\
Jeffrey Carver &  University of Alabama\\
Emily Chen & NCSA, University of Illinois\\
Sou-Cheng Choi & NORC at the University of Chicago \&  Illinois Institute of Technology\\
Yuhan Ding &  Illinois Institute of Technology\\
Lorraine Hwang &  CIG,  UC Davis\\
Ray Idaszak &  RENCI, University of North Carolina at Chapel Hill\\
Frank L\"{o}ffler &  Louisiana State University\\
Abigail Cabunoc Mayes &  Mozilla Science Lab\\
Pate Motter &  University of Colorado\\
Kyle Niemeyer &  Oregon State University\\
Birgit Penzenstadler &  California State University Long Beach\\
Bernadette Randles &  UCLA\\
Ilian Todorov &  STFC Daresbury Laboratory\\
Nic Weber &  University of Washington iSchool

\end{longtable}
}

\section{Best Practices Group Discussion}
\label{sec:appendix_best_practices}

Sandra Gesing\footnote{email:
\href{mailto:sandra.gesing@nd.edu}{sandra.gesing@nd.edu}} will serve as the
point of contact for this working group, and be responsible for ensuring timely
progress of the planned actions.

\subsection{Group Members}

\begin{itemize}
\item Abani Patra -- University at Buffalo
\item Sandra Gesing -- University of Notre Dame
\item Neil Chue Hong -- Software Sustainability Institute
\item Gregory Tucker -- University of Colorado at Boulder
\item Birgit Penzenstadler -- California State University Long Beach
\item Abigail Cabunoc Mayes -- Mozilla Foundation
\item Frank L\"{o}ffler -- Louisiana State University 
\item Colin C. Venters --  University of Huddersfield
\item Lorraine Hwang -- UC Davis 
\item Sou-Cheng Choi -- NORC at the University of Chicago \&  Illinois Institute of Technology
\item Suresh Marru -- Indiana University
\item Don Middleton -- NCAR 
\item Daniel S. Katz --  University of Chicago \& Argonne National Laboratory
\item Kyle Niemeyer -- Oregon State University
\item Jeffrey Carver -- University of Alabama
\item Dan Gunter -- LBNL
\item Alexander Konovalov -- University of St Andrews
\item Tom Crick --  Cardiff Metropolitan University

\end{itemize}

\subsection{Summary of Discussion}

Core questions that will need to be explored are in reliability,
reproducibility, usability, extensibility, knowledge management, and continuity
(transitions between people). Answers to these will guide us on how a software
tool becomes part of the core workflow of well identified users (stakeholders)
relating to tool success and hence sustainability.
Ideas that may need to be explored include:
\begin{itemize}

\item Requirements engineering to create tools with immediate uptake;

\item When should software ``die''?

\item Catering to disruptive developments in environment, e.g., new hardware,
new methodology;

\item Dimensions of sustainability: economic, technical, environmental,
and obsolescence.

\end{itemize}

Sustainability requires community participation in code development and/or a
wide adoption of software. The larger the community base is using a piece of
software, the better are the funding possibilities and thus also the
sustainability options. Additionally, the developers’ commitment to an application is
essential and experience shows that software packages with an evangelist
imposing strong inspiration and discipline are more likely to achieve
sustainability. While a single person can push sustainability to a certain
level, open source software also needs sustained commitment from the developer
community. Such sustained commitments include diverse tasks and roles, which can
be fulfilled by diverse developers with different knowledge levels. Besides
developing software and appropriate software management with measures for
extensibility and scalability of the software, active (expertise) support for
users via a user forum with a quick turnaround is crucial. The barrier to entry
for the community as users as well as developers has to be as low as possible.

\subsection{Description of Opportunity, Challenges, and Obstacles}

There is an opportunity to collaborate on a white paper, which will be revisited
regularly for further improvements, to enhance knowledge of the state of best
practices, resulting in a peer-reviewed paper. We would like to reach a wide
community by doing this. But these are also the challenges and obstacles -- to
get everyone to contribute to the paper and to reach the community.

\medskip
\noindent{\bf White Paper Outline:}
\begin{enumerate}
\item Introduction and Scope of White Paper 
\item Related Work
\item Case Studies
\begin{enumerate} 
\item PETSc
\item NWChem
\item CIG
\end{enumerate}
\item Community Related Practices
\begin{enumerate} 
\item Findings
\item Recommendations
\end{enumerate}
\item Governance and management
\begin{enumerate} 
\item Findings
\item Recommendations
\end{enumerate}
\item Funding Related
\begin{enumerate} 
\item Findings
\item Recommendations
\end{enumerate}
\item Metrics for sustainability
\item Tools
\item Conclusions
\end{enumerate}

\subsection{Key Next Steps}

The key next steps are to write an introduction, reach out to the co-authors,
and to agree on the scope of the white paper.

\subsection{Plan for Future Organization}

Sandra Gesing and Abani Patra are the main editors and will organize the overall
communication and the paper. Sections will be assigned to diverse co-authors.

\subsection{What Else is Needed?}

At the moment we do not see any further requirements.

\subsection{Key Milestones and Responsible Parties}
\begin{itemize}
\item 28 Dec: Introduction and scope finished (Abani Patra/Sandra Gesing)
\item 06 Jan: Sections assigned (Abani Patra/Sandra Gesing)
\item 31 Jan: Analyzing funding possibilities for survey
\item 31 Jan: First version of each section
\item 15 Feb: Distribution to the WSSSPE community
\item 31 Mar: Final version of the white paper
\item 30 Apr: Submission to a peer-reviewed journal?
\end{itemize}

\subsection{Description of Funding Needed}
We might need funding for a journal publication (open-access options).

\section{Funding Research Programmer Expertise Group Discussion}
\label{sec:appendix_funding_spec_expert}

James Hetherington\footnote{email:
\href{mailto:j.hetherington@ucl.ac.uk}{j.hetherington@ucl.ac.uk}} will serve as
the point of contact for this working group, and be responsible for ensuring
timely progress of the planned actions.

\subsection{Group Members}

The group at WSSSPE:

\begin{itemize}
\item Don Middleton -- National Center for Atmospheric Research
\item Joshua Greenberg -- Alfred P. Sloan Foundation
\item James Hetherington -- University College London
\item Lindsay Powers -- The HDF Group
\item Mark A. Miller -- San Diego Supercomputer Center
\item Dan Sellars -- CANARIE
\end{itemize}

This was further enhanced by additional discussions at the following
GCE15 conference:
  
\begin{itemize}
\item Lorraine Hwang -- UC Davis
\item Simon Trigger -- BioTeam, Inc.
\item Nancy Wilkins-Diehr -- San Diego Supercomputer Center
\item Alexander Vyushkov -- University of Notre Dame
\item Sandra Gesing -- University of Notre Dame
\item Ali Swanson -- University of Oxford

\end{itemize}

\subsection{Summary of Discussion}

In addition to the points noted in the main discussion (\S\ref{RSE}), we also
discussed the following:

``Are you an RSE or a RA?'' is not properly a binary question. Most of us sit at
different points on that spectrum, and move along it during our careers (usually
from RA to RSE---examples of movement in the other direction from readers would
be welcomed). Either way, the label ``Research Software Engineer'' is now
starting to have some power. Many scientists do not want to be writing code;
some do, to varying degrees. These groups can usefully support each other.

What is the power of the label? How can we get the word out about RSE support
using the label?

Will research science developers be required in the long run? One issue that
came up was whether the need for developers was a time bounded one; is it the
case that the new generation of computer and software savvy scientists will be
so comfortable in developing their own code that the professional developers
will not be needed? And this brings up the flip side question, ``Do scientists
really want to be writing code?''

We also had a little discussion about how to make a career path for research
developers. It need not be solely an academic enterprise, but in the past tenure
has often been problematic for people of this class.

Skills and resources may vary between teams. To help resolve this, maintaining
high levels of communication between groups will be valuable. In the United
Kingdom (UK), there are plans to permit resource sharing between institutional
RSE groups. Perhaps there are circumstances under which an RSE skill exchange
could be arranged, either formally or informally.

Collaborative funding can be crucial to RSE groups, to ensure that research
leadership remains with the domain scientists. As an example, at NCAR, university partnerships
are required for submission of proposals, so collaboration is an essential part
of grant submission, and this will tend to bring developers and scientists
together. The UCL group also follows this approach, with all bids requiring an
academic collaborator.

Domain scientists and developers are funded together in a single proposal.
Another example of a success is the development of semantics and linked data in
support of ocean sciences. An EarthCube-funded project pairs domain scientists
with RSEs and has been successful; the semantics attached have increased data
use and discovery significantly.

An alternative approach has been the provision of programming expertise as part
of national compute services. The US XSEDE project's Extended Collaboration Support 
Services (ECSS) is a set of developers who are paid with XSEDE funding, and are
on ``permanent'' staff. When PIs request allocations on XSEDE resources, there
is a finite pool of developer time that can be awarded, typically for one year
only, and at partial effort, typically 20 percent or so. The finite time allowed
provides motivation for the scientist and the scientist's group to work closely with the
developer and to become educated in what the developer is doing, so they can
sustain the effort once the ECSS period is over. This funding mechanism can be
highly efficient for scientific problems, because the developer pool assembled
by the research providers are, by definition, expert in the characteristics of
their specific resource, and can very quickly assess the scientist's needs, and
what it will take to implement software that meets the user's needs. However, it
does not develop capacity within institutions, and since XSEDE is a time-bounded
program, it should not be relied upon as a long-term solution to acquiring this
type of capacity.

The UK allows this kind of collaboration to support the creation of scientific
software for the large supercomputing resource (ARCHER). However, while the
support can come from the staff of the Edinburgh Parallel Computing Centre,
who hosts the computer, this ``embedded CSE'' resource also funds the
programming coming from local groups. This has been very helpful in providing
funding to establish local groups. These groups work best when they develop good
collaborations with national cyberinfrastructure pools. When an organization
assembles a developer pool, diversity is developed and skills can be
transferred.

We would like to see these models applied outside high performance computing.
Most scientific software is not destined to run on national cyberinfrastructure,
but needs similar support. The argument regarding making better use of expensive
hardware through software improvements has been useful politically, (and many
RSE groups are cited in organizations which host clusters for this reason), but
the time has come to make the case that software itself is a critical
cyberinfrastructure, and, with a much longer shelf-life than hardware, is itself
a capital investment.

The CANARIE group (Canada) accepts proposals for providing services to broad
communities, integrating people who are doing things that are complementary. The
goal is to make the available stack more robust and richer for everyone. They
offer short cycles of funding for creating some useful functionality that shows
a diversity of input and draws from across disciplines as a key metric, If this
metric is met successfully, then more funding may follow. This can apply within
or across institutions.

There can be problems communicating across cultural barriers, with domain
scientists seeing developers as ``other''. Both collaboration and tools to fund,
encourage, or motivate collaboration are extremely important. 

We think support from non-governmental organizations will be important if RSE
groups will become established. The Sloan Foundation is currently funding data
science engineers, who work in the context of other software developers at the
University of Washington. These scientists work in the e-Science Studio/Data
Science Studio, and they help a group of graduate students in solving their
problems in data science and data management. During Fall and Spring, a 10-week
incubator program allows students to work two days a week on a data-intensive
science project. Some fraction of the developer time is dedicated to the
developers' personal interests as well as instruction.

The goal for Sloan is to obtain success stories and demonstrable value
of the presence of data scientists on university staff. These stories are the
basis for arguments to the host organization. This is an effort to create
awareness of the value of research scientist developers. Embedding with
scientists, and adding spare capacity is critical to making the innovation
possible. This model is essentially to argue for permanent budget lines to
support data scientists as part of university staff hires, just as with core
facilities. This could become a fee-for-service model requested by grant
funding, just as DNA sequencing is for core facilities, if it becomes apparent
that this gives competitive advantage to a university's research effort.

One model that has been helpful in finding funding for RSE groups is the use of
funds left over on research grants when RAs have left prematurely -- PIs like
this arrangement as it is hard to find good staff for short-term positions, so
having a pool of research programming staff on hand resolves this problem. We
recommend that funders give explicit guidance to grant holders and institutions
that such an arrangement is favorable. Framework agreements permitting this to go
ahead without checking back every time with funders and/or grant panels would
further smooth this. (This also provides more stable jobs for those who hold
these skills, but arguments about making life nicer for postdocs will not help
persuade funders or PIs!)

There is some question about the most effective duration and percentage of full
time for a programmer's work on a project. At least three months is necessary
for the programmer to read into the science (RSEs must not become so disengaged
from research that they do not have time to read a few papers -- this will result
in code which does not meet scientific needs), but too long could result in an
RSE losing their flexibility, becoming so engaged in one project that when that
project ends, they find it hard to transfer. For this reason, we recommend
that 40\% is ideal; two projects per developer, with some time for
training and infrastructure work. Having two developers per project seems to be ideal,
in the sense that software development is enhanced by two pairs of eyes.

There is, as yet, no clear answer as to the scale of aggregation needed to make
such a program work. A university wide program allows enough scale to be robust
to fluctuations of funding within one field. But a specialization focus on
developers to support, for example, physical or biological sciences may be
preferable, if the customer base is large enough. The desire to aggregate enough
work to make it sustainable, and the need to have domain-relevant research
programming skills, are in tension.

In the UK, another source of funding for research software is the
Collaborative Computational Projects (CCPs): domain specific communities put
forward proposals that are a priority of the community as a whole, for example,
biosimulation or plasma physics. These bodies act as custodians of community
codes, and a central team also provides software engineering support.

However this area develops, the need for funding for software as a
cyberinfrastructure component is clear.  Funding that permits code to be
refactored, tidied, and optimized is rare; this is often done ``on the sly'' in a
scientifically focused grant. The UK EPSRC's ``software for the future'' call,
which really permits explicit investment in software as an infrastructure, is so
oversubscribed as to have a 4\% success rate; the demand is clear!

One opportunity is the idea of co-design, where infrastructural libraries are
developed alongside the scientific codes that will call them. However,
collaboration is hard to foster here; as incentive structures are still focused
on short-term papers. This can cause infrastructure developers to focus more on
publications in their areas of mathematics and computer science, the domain
developers on the shorter-term needs of their own fields. Genuine collaborative
co-construction is harder to foster.

It can be more difficult to help leading domain scientists see the value of
engineering effort than those in their teams who are forced to work with
difficult-to-use or unreliable software tools, as they do not see the pain.
Perhaps a version of ``software carpentry'' targeted at those PIs who are
awarded or apply for software-intensive grants could be valuable here.

RSEs provide a useful contribution to their universities' teaching missions, as
well as research, as they are well placed to deliver the research programming
training that many scientists now need. In the longer term, with programming
skills taught to all through their careers, we hope specialist scientific
developers will be less needed.

\subsection{Key Next Steps}

We will seek to identify and approach existing research programming organizations,
to get their permission to list them on a list of research software groups.
Casual conversation during the meeting made it clear that although the title is
not widely used in the US, this position is not rare. We spoke with several
individuals who, at distinct universities, had RSEs (in effect if not in name)
who were funded under differing models.

We will also look for examples of groups which have successfully become self-
sustaining following initial seed funding.

In this respect, information gathering via a survey and subsequent analysis could be
very useful. We would need to assemble a list of targeted individuals. (What
positions and ranks are likely to know and care enough to respond?) Perhaps the
Science Gateway Institute has already acquired information that could be helpful
to advance this issue, and/or craft a proper survey and suggest target individuals.

\subsection{Plan for Future Organization and Future Needs}

The UK RSE community will provide initial facilities to host this list, and
continue to work to spread the initiative, but local leadership in the US is
needed if this campaign is to succeed. This will require an initial gathering of
identified research software organizations in the US to this end.

\subsection{Description of Funding Needed}

Financial support for an initial conference that brings together research software
groups to form an organization and create a resource sharing structure would
help to further this campaign. Funding to conduct and analyze a survey could
also be quite useful as knowing where we stand today, and what models are in use
could fuel the ideas for further development of developers in this category.

In the longer term, funding organizations, especially non-governmental
organizations with the capability to effect innovation through seed funding,
could provide support to nucleate the creation of research software groups. As
noted above, Sloan has already initiated one such program, and collaboration
with Sloan or at least study of their methods and success or failure could be
extremely useful in approaching universities and other institutions in funding
this development track. It seems clear that if the value proposition can be made
to university administrators, this track could flourish with buy-in at the
administrative level.

\section{Transition Pathways to Sustainable Software: Industry \& Academic Collaboration Working Group Discussion}
\label{sec:appendix_industry_interaction}

Nic Weber\footnote{email: \href{mailto:nmweber@uw.edu}{nmweber@uw.edu}} will
serve as the point of contact for this working group.

\subsection{Group Members}

\begin{itemize}
\item Nic Weber -- University of Washington
\item Suresh Marru -- Indiana University
\item Jeffrey Carver -- University of Alabama
\item Davide DelVento -- NCAR/CISL
\item Steven Brandt -- Louisiana State University
\end{itemize} 

\subsection{Summary of Discussion}

The group's initial broad question was, ``What makes for successful transitions of
scientific software from academia to industry?'' There are a number of potential
funding transitions that may occur:
\begin{itemize}

\item A project could be \textbf{refunded}, and development or maintenance of
the software continue as planned.

\item A project might locate a \textbf{new source of funding} in which case the
software may be further developed or simply maintained as before.

\item The project could transition to a \textbf{community supported model}
whereby the software's ownership, maintenance, and stewardship  become similar
to peer-production models in open-source (e.g., see
Howison~\cite{howison_sustaining_2015}).

\item The project could receive some form of industry sponsorship in which case
ownership of the intellectual property, licensing, maintenance activities,
hosting, etc.\ may change significantly.

\item The project could gain attention from a industry use case who would
potentially make in-kind contributions by having paid staff contribute to the
software.

\end{itemize}

We characterized each of the above potential changes in funding as ``transition
pathways'' to sustainable software (see similar work by Geels and
Schot~\cite{Geels:2007}).

Our work at WSSSPE3 included the following three activities (described in more detail
below): (1) brainstorming goals for this type of research, (2) imagining
potential outcomes of completing a set of case studies on this topic, and (3)
generating a set of working definitions for some of the broad concepts we are
describing.

First, we discussed the \textbf{goals} of this research, attempting to answer the 
question \emph{What is the goal of doing research on transition pathways?}
A number of research questions arose:  Can we identify collaborations that have 
occurred and try to understand which were successful, which were unsuccessful, 
and what factors contributed to these successes/failures? Can we determine what 
each partner wants to get out of such a collaboration? For example, why would 
industry be interested in collaborating with academia? Or why would academia 
be interested in collaborating with industry? How could we design a study that 
focused on the impact of the software in undergoing this type of transition?

Next, we imagined \textbf{potential outcomes} of research on this topic, involving 
a set of case studies that look at successful and unsuccessful
transitions of researchers between academia and industry. This might address 
each of the transition types (as described below). Successful transitions are
described as those that lead to either weak or strong sustainability (also
defined below). In addition, the results from this research might help create a 
generalizable framework that might allow for the study of different transition 
pathways (other than academia to industry).

Finally, we created some \textbf{general definitions} for these concepts; we 
characterize transitions in the following ways:
\begin{itemize}

\item Handoff model: academia initially writes the software, industry (for-profit 
or nonprofit) then takes over the project.

\item Co-Production Model: industry and academia interact throughout development
of the project.

\item Sponsorship Model: academia writes and maintains the software; 
industry contributes funding for the development\slash maintenance of software.
In this example, industry is also likely a user of the software.

\item Spinoff model: transition to a for-profit or non-profit company owned by or in
collaboration with original developers.

\end{itemize}

We characterized sustainability in the following ways:
\begin{itemize}

\item Weak Sustainability: Software continues to be accessible, useful, and
usable.

\item Strong Sustainability: Software meets criteria above, but is also able to
be reused for further innovation (i.e., issued non-restrictive open-source
license).

\end{itemize}
We refer readers to Becker et al.~\cite{Becker:2014} for an extended discussion
of weak versus strong sustainability.

\subsection{Description of Opportunity, Challenges, and Obstacles} 

The opportunity is to create a catalog of success/failure for current and future
software projects to be prepared for transitions and achieve sustainability of
the software.

The obstacle is more superficial, in finding a champion to gather such
information. It will be a challenge to keep this information and surveys
updated. With changing rapidly changing industry landscapes, an obsolete survey
could be of less or no use.

\subsection{Key Next Steps}

Identify projects that are collaborative, perhaps by reviewing funded projects
from programs specifically geared towards industry academic collaborations.

Develop a systematic process for conducting case studies (what kind of data are
being gathered about each case)

\subsection{Plan for Future Organization}

No concrete plans have been made at this point. If the community can rally
behind this topic, some momentum could be built. Those interested should post at
\url{https://github.com/WSSSPE/meetings/issues/46}

\subsection{What Else is Needed?}
Nothing at the moment. 

\subsection{Key Milestones and Responsible Parties}

A key portion of this effort will require focused surveys of projects which have
succeeded and failed in transition. Both these categories will yield good
learning on what works and what does not work. The group has identified what
needs to be studied further, but has not identified responsible parties to
conduct them.

Community members could help in gathering data by means of interviews, historical
documents or documentation, and surveys.

An example of data collection is: 
\begin{itemize}
\item Origin: Where did project start? 
\item People involved: How many people in original project were involved in
transition/collaboration?
\item Specs on software
\item Language
\item Size
\item Hardness (age)
\item Lead-up to Transition: How long was project in development before it began transition?
\item Motivation for Transition: Why was transition initiated? By whom? 
\end{itemize}

\subsection{Description of Funding Needed}

Concrete funding needs were not discussed in this working group but a general
impression was that some seed funding would motivate members of this group or
others in community to launch a survey effort.

\section{Engineering Design Group Discussion}
\label{sec:appendix_eng_design}

Birgit Penzenstadler\footnote{email:
\href{mailto:birgit.penzenstadler@csulb.edu}{birgit.penzenstadler@csulb.edu}}
and Colin C.\ Venters\footnote{email:
\href{mailto:c.venters@hud.ac.uk}{c.venters@hud.ac.uk}} will serve as the points
of contact for this working group, and be responsible for ensuring timely
progress of the planned actions.

\subsection{Group Members}

\begin{itemize}
\item Birgit Penzenstadler -- California State University, CA, USA
\item Colin C.\ Venters -- University of Huddersfield, Huddersfield, UK
\item Matthias Bussonnier -- UC Berkeley, CA, USA
\item Jeff McWhirter -- Geode Systems 
\item Patrick Nichols -- National Center for Atmospheric Research, CO, USA
\item Ilian Todorov -- Science \& Technology Facilities Council, UK
\item Ian Taylor -- Cardiff University, UK
\item Alexander Vyushkov -- University of Notre Dame, IN, USA
\end{itemize}

\subsection{Summary of Discussion}

This group was comprised  of members from different backgrounds, including quantum
chemistry, epidemiology, microscopy, computer science, and software engineering.
Each participant was invited to give their perspective on the topic area and
what they thought were the crucial points for discussion. There was a general
consensus that there was a need for relating principles to practice for the
computational science and engineering community. Furthermore, various members of
the group expressed their interest in tools and best practices for facilitating
the maintenance and evolution of scientific software systems. It was agreed to
identify principles from software engineering and from sustainability design
and, based on those lists, discuss what each of those would mean applied to
specific example systems from the expert domains of some of the group members.
The group identified a number of software engineering principles drawn from the
Software Engineering Body of Knowledge (SWEBOK)~\cite{swebokv3}.

Software design principles include abstraction, coupling and cohesion,
decomposition and modularization, encapsulation and information hiding,
separation of interface and implementation, sufficiency completeness and
primitiveness, and separation of concerns. Similarly, user interface design
principles include learnability, user familiarity, consistency, minimal
surprise, recoverability, user guidance, and user diversity. The sustainability
design principles were drawn from the Karlskrona Manifesto on Sustainability
Design~\cite{Becker:2014}. The manifesto states that sustainability is systemic,
multidimensional, and interdisciplinary; transcends the system's purpose; applies to
both a system and its wider contexts; requires action on multiple levels;
requires multiple timescales; changing design to take into account long-term
effects does not automatically imply sacrifices; system visibility is a
precondition for and enabler of sustainability design. A number of sustainable
software engineering principles proposed by Tate~\cite{tate:2005} were also
considered including: continual refinement of product and project practices; a
working product at all times; continual emphasis on design; and value defect
prevention over defect detection.

This congregated list is an initial collection of principles that could be
extended by adding from further related work form separate disciplines within
the field of software engineering, including requirements engineering, software
architecture, and testing. The group identified two example systems to discuss
the application of the principles. The first one was a quantum chemistry system
that allows the analysis of the characteristics and capabilities of molecules
and solids. The second one was a modeling system for malaria that permitted
biologists to analyze a range of datasets across geography, biology, and
epidemiology, and add their own datasets. The group then examined the principles
and took a retrospective analysis of what the developers did in practice against
how the principles could have made a difference. This raised the question, what
do the principles mean for computational scientific and engineering software?
Similarly, how do the principles relate to non-functional requirements? It was
suggested that at the very minimum, that sustainable software engineering
principles should be mapped to two core quality attributes that underpin
technically sustainable software:
\begin{itemize}
\item Extensibility: the software's ability to be extended and the level of
effort required to implement the extension;

\item Maintainability: the effort required to locate and fix an error in
operational software.

\end{itemize}
These fundamental building blocks could then be extended to include other
quality attributes such as portability, reusability, scalability, usability, and
energy efficiency etc. Nevertheless, this raises the question of what metrics
and measures are suitable to demonstrate the sustainability of the software. In
addition, what do the five dimensions of sustainability mean for scientific
software, i.e., environmental, economic, social, technical and individual?

\subsection{Description of Opportunity, Challenges, and Obstacles}
The opportunity was identified to distill existing software engineering and
sustainability design knowledge into ``bite sized'' chunks for the Computational
Science and Engineering Community. In addition, two primary challenges were
identified:
\begin{itemize}
\item Mapping of the principles to best practices.
\item Demonstrating the return on investment of those best practices.
\end{itemize}

\subsection{Key Next Steps}

In order to achieve the following three goals: (1) a systematic analysis of a number of example systems from
different scientific domains with regards to the identified principles, (2) the
identification of the commonalities and gaps in applying the principles to different
scientific systems, and (3) a proposal of a set of guidelines on the
principles, the following next steps were discussed.

\subsection{Plan for Future Organization}

The following plan for future organization was discussed:
\begin{itemize}
\item Identify suitable undergraduate or post-graduate students.
\item Design and pilot study.
\item Organize coordinating online calls via Google Hangout.
\end{itemize}

\subsection{What Else is Needed?}

\begin{itemize}
\item Ethics committee review panel approval required for data collection.
\end{itemize}

\subsection{Key Milestones and Responsible Parties}

The following key milestones were discussed as a roadmap for the set of
guidelines on software engineering principles:
\begin{itemize}
\item Oct/Nov 2015: Study design and interview guideline
\item Jan/Feb 2016: Interviews conducted and transcribed
\item Mar/Apr 2016: Analysis complete
\item May 2016: Report written
\end{itemize}

\subsection{Description of Funding Needed}

Specific funding was not discussed in this working group. However, this is a
open topic that can be discussed in relation to emerging funding calls from
National agencies or grant proposal initiatives.

\section{Metrics Working Group Discussion}
\label{sec:appendix_metrics}

Gabrielle Allen\footnote{email: \href{mailto:gdallen@illinois.edu}{gdallen@illinois.edu}} 
will serve as the point of contact for this working group.

\subsection{Group Members}

\begin{itemize}
\item Gabrielle Allen -- University of Illinois at Urbana-Champaign
\item Emily Chen -- University of Illinois at Urbana-Champaign
\item Neil Chue Hong -- U.K. Software Sustainability Institute
\item Ray Idaszak -- RENCI, University of North Carolina at Chapel Hill
\item Iain Larmou -- Engineering and Physical Sciences Research Council
\item Bernie Randles -- University of California, Los Angeles
\item Dan Sellars -- Canarie
\item Fraser Watson -- National Solar Observatory
\end{itemize}

\subsection{Summary of Discussion}

The group discussion began by agreeing on the common purpose of creating a set
of guidance giving examples of specific metrics for the success of scientific
software in use, why they were chosen, what they are useful to measure, and any
challenges and pitfalls; then publish this as a white paper. The group discussed
many questions related to useful metrics for scientific software as follows:
\begin{itemize}

\item
Is there a common set of metrics, that can be filtered in some way

\begin{itemize}
\item
        Does this create a large cost
\end{itemize}

\item
Can we fit metrics into a common template (i.e., for collection, for description)

\item
Which would be the most useful ones

\begin{itemize}
\item
        Which ones would be most useful for each stakeholder
\end{itemize}

\item
Which ones are the most helpful, and how would we assess this

\item
How do you monitor

\begin{itemize}
\item
        Self-checking - if monitoring is done in the open, then people will call out cheats
\end{itemize}

\item
Should this be published with the software metadata

\begin{itemize}
\item
        This would make it easier for public to see the metadata

\item
However, there is no commonly-used standard (DOAP is a good standard but not
widely adopted)

\item
        The Open Directory Project (ODP) metadata is available for UK infrastructure
\end{itemize}

\item
Intersection of most useful and easiest to collect should be explored

\item
How can students/curricula be used as part of a solution

\item
Number of users could be affected by other metrics including, e.g., accessibility

\item
Assume metrics are collected properly, but guidance should be provided none-the-less

\item
Continuum for each metric

\begin{itemize}
\item
Ideal situation is the absolute minimum, so that people can decide on their own
what the cost versus usefulness tipping point is
        
\end{itemize}

\item
Maturity plays a part

\begin{itemize}
\item
        Consider different metrics brackets for different maturity levels
\end{itemize}

\item
What are we using metrics for

\begin{itemize}
\item
        What software should I use if I have a choice

\item
Where should funders place funding for best impact (e.g., funding two-star
software versus three-star) and where there are gaps

\item
        How to promote reduction of code proliferation

\item
        Metrics used for software panels to provide information

\item
        Metrics used for finding problems in their systems

\end{itemize}

\item
Can we use metrics to help people identify the best codes as part of a community
effort

\end{itemize}

\smallskip
\noindent
Next, a roadmap for how to proceed was discussed including creating a set of
milestones and tasks as follows:
\begin{itemize}
\item
Can we create a roadmap and milestones for this activity

\item
Need to come up with a set of tasks

\item
Go to NSF Software Infrastructure for Sustained Innovation (SI2) projects asking
them what metrics they defined, and how useful they were

\begin{itemize}
\item
        Milestone: Create report which assesses the metrics that SI2 projects used

\begin{itemize}
\item
Ask SI2 PIs to say what metrics they said they would use (copied from proposal)
\item
                Ask SI2 PIs what numbers they reported

\item
                Ask SI2 PIs what they would have changed

\item
                A UIUC student on the project will work on this
\end{itemize}

\item
        Tentatively aim for March 2016
\end{itemize}

\item
Do something similar for UK SFTF and TRDF software projects to ask them what
would be useful metrics to report; also eCSE projects

\begin{itemize}
\item
        Compare these to understand if there were any implications for including metrics
\end{itemize}

\item
Collaboratively create plan and documentation for doing this

\begin{itemize}
\item
Give some examples from group members projects, and aim to build out some of the
measurement continuum

\item
        Road-test at the WSSSPE4 meeting
\end{itemize}

\item
Collect the various frameworks together and do a comparison summary

\end{itemize}

\smallskip
\noindent
The idea was put forth for the group to interact with the organizing committee
of the 2016 NSF Software Infrastructure for Sustained Innovation (SI2) PI
workshop in order to email out a software metrics survey to all SI2 and related
awardees as a targeted and relevant set of stakeholders. This survey would be
created by one of the student group members. Similarly, it was suggested that a
software metrics survey be sent to the UK SFTF and TRDF software projects to ask
them what metrics would be useful to report. The remainder of the discussion
focused mainly on the creation of a white paper on this topic. This resulted in
a paper outline and writing assignments with the goal of publishing in venues
including WSSSPE4, IEEE CISE, or JORS.

\subsection{Description of Opportunity, Challenges, and Obstacles}

The following opportunities, challenges, and obstacles were discussed:
\begin{itemize}
\item
Metrics are important for:

\begin{itemize}
\item
        Tenure and promotion

\item
        Scientific impact

\item
        Discovery

\item
        Reducing duplication

\item
        Basis for potential industrial interest in adopting software

\item
        Make case for funding
\end{itemize}

\item
No commonly used standard for collecting or presenting metrics

\item
We do not know if there is a common set of metrics

\item
We have to persuade projects that it is useful to collect metrics

\end{itemize}

\subsection{Key Next Steps}

The following next steps were discussed:
\begin{itemize}
\item
Skype phone call to coordinate shortly after the conclusion of the WSSSPE3 workshop

\item
Get started on IRB at University of Illinois Urbana-Champaign in anticipation of
SI2 project survey (may need more thought into survey)

\item
Get started on white paper and associated survey

\end{itemize}

\subsection{Plan for Future Organization}

The following plan for future organization was discussed:
\begin{itemize}
\item
Our group has created a white paper outline with sections assigned to the above
individuals; see timeline below. 

\item
Organize coordinating phone calls.

\end{itemize}

\subsection{What Else is Needed?}

The following list of what else is needed was discussed:

\begin{itemize}
\item
IRB approval/exemption needed for surveys, collecting data

\item
Coordination with 2016 NSF SI2 PI workshop organizing committee to possibly
piggyback on this event to offer survey to attendees in advance

\item
Coordination (mail communication, info page, etc.) via WSSSPE GitHub or other
means?

\end{itemize}

\subsection{Key Milestones and Responsible Parties}

The following items were discussed as a roadmap for the production of a white
paper:
\begin{enumerate}
\item
October -- November 2015: IRB paperwork as appropriate completed (Gabrielle
Allen and Emily Chen)

\item
October -- December 2015: Draft white paper sections 1-3 (the paper outline has
initial writing assignments)

\item
October -- December 2015: Run surveys and collect information

\begin{enumerate}
\item
Piggyback on planning for 2016 NSF SI2 PIs meeting to be held Feb 16-17, 2016 
       
\end{enumerate}

\item
January -- February 2016: Analyze results of data collection from projects

\item
March -- April 2016: Draft sections 4-7 of the white paper

\item
May 2016: Draft section 8-9 of the white paper

\item
May -- June 2016: Get initial feedback from members of the community and revise

\item
Est.\ July 2016: By the time of next CFP for WSSSPE, to have a complete draft of
the white paper

\item
Est.\ Sept -- Oct 2016: Responses to white paper to be submitted to WSSSPE4

\end{enumerate}

\subsection{Description of Funding Needed}

Funding needs were not discussed in this working group and it was thought that
this could potentially be revisited down the road.

\section{Training Working Group Discussion}
\label{sec:appendix_training}

Nick Jones\footnote{email:
\href{mailto:nick.jones@nesi.org.nz}{nick.jones@nesi.org.nz}} will serve as the
point of contact for this working group, and be responsible for ensuring timely
progress of the planned actions.

\subsection{Group Members}
\begin{itemize}
\item Nick Jones -- New Zealand eScience Infrastructure
\item Iain Larmour -- Engineering \& Physical Sciences Research Council, UK
\item Erin Robinson -- Foundation for Earth Science
\end{itemize}

\subsection{Summary of Discussion}

While little training focuses specifically on sustainable
software, a variety of training activities could increase researcher awareness of
and engagement with software professionals and software engineering practices.
Research Software Engineers are being recognized as critical contributors to
high quality research; the pathway to acquire and master the relevant skills
is not yet clear; equally those skills required by researchers in general are
also not commonly understood nor routinely developed.

The group's discussion explored a rapidly growing array of training that is seen
to contribute to sustainable software. The offerings are diverse, including:
self-paced online modules focused around specific tools; single and multiple day
training workshops that raise awareness of a tool chain to support collaborative
and shared software development within a research workflow; block courses
specializing on particular methods, technologies, and applications; academic
programs at undergraduate and masters levels; doctoral training programs that in
part contain requisite skills training activities.

While some of this training focuses on applying software engineering practices
within the context of research, meeting the values and goals of research are
less often incorporated as explicit learning outcomes. With software (and
similarly, data) often being the only tangible artifact of a research method or
protocol, the dependency between software applications and the quality of
research adds complexity to the learner's journey. In recognition of the longer
term investment required by researchers to integrate such skills into their
research practices, many activities are focusing on emotionally engaging
researchers and cohorts, to build a sense of shared purpose beyond the obvious
goal of technical skill acquisition.

In reviewing current training activities, the group identified a variety of
perspectives seen as useful in positioning activities in ways to better
communicate why and when best to apply each activity. Training can be
categorized on a variety of spectra, with content and delivery ranging across them, for example:
programming to research; basic to advanced; technical to emotional; informal to
formal; and self-paced to participative. A few attempts have been made to situate a
cross section of training activities within such dimensions, creating easier
means of communicating the value of any specific opportunity and the pathways
across opportunities over time.

Evaluation of training delivery and outcomes is seen as a weakness common to
most non-academic training activities. Opportunities for measuring success in
delivering training start simply with collecting a Net Promoter Score, which
lets those delivering training know whether attendees are likely to recommend
the training to others. In looking at the longer term outcomes for the learner,
frameworks such as Bloom's taxonomy and Kirkpatrick's evaluation model offer possible
approaches.

In this latter case of formal evaluation, ownership of evaluation as a component
of career development for any researcher appears mostly absent. While academic
research institutions have professional development centers to support research
staff, the skills taught which might impact on sustainable software are limited
at best, and lack a clear and coherent development pathway.

Coordination of these training projects will depend on buy-in from a broad range
of training program and activity leaders, suggesting a key opportunity lies in
identifying and bringing together these people on a regular basis.

\subsection{Description of Opportunity, Challenges, and Obstacles}

Software skills are needed by an increasing array of researchers and fields. The
training arc is not well-defined, with a sometimes baffling array of training
opportunities responding to various facets of skill deficit and need. Given this
current complexity, coordination across training projects would create common
frames of reference, communicating and integrating activities to better serve
the needs of researchers.

Building this community could lift the maturity of training projects and
capabilities, enabling more advanced approaches to address key gaps in
evaluation, career development, and a lift in the standard of research
practices.

In aiming at these opportunities, it will be necessary to find the means to
support those involved in leading training activities to allocate time to
coordination activities, which will often sit beyond their current scope of
responsibility.

These activities are also distributed globally, with no single country or region
offering a comprehensive set of capabilities and initiatives. Any coordination
activity will therefore need to raise the profile of the opportunity gap with
relevant research funders and policy makers.

\subsection{Key Next Steps}

The goal of the following next steps is to quickly test whether there is
interest in establishing a community committed to increasing the degree of
coordination across training projects.

\begin{enumerate}

\item Hold a virtual meeting by December 2015, to bring together a broader group
of interest in this topic, with specific goals to:

	\begin{enumerate}
	    
	\item Identify programs with existing activities aimed at integrating across
	training projects.
	        
	\item Identify training projects with an interest in participating in
	coordination efforts.
	        
	\item Identify funding opportunities to bring together training program and
	project leaders to identify shared goals for future coordination of activities.
	        
	\item Agree on a communications plan to qualify whether programs, projects, and
	funders are interested in engaging and committing to ongoing activities.
	        
	\end{enumerate}
    
\item Review progress within 3 months, to establish next steps, if any.

\end{enumerate}

\subsection{Plan for Future Organization}

Continue to track progress by posting comments to WSSSPE3 issue.

\subsection{What Else is Needed?}

If the group moves from early-stage formation into working towards shared goals,
expertise will likely be required in pedagogy and training evaluation.

\subsection{Key Milestones and Responsible Parties}
\begin{enumerate}

\item October through December, Nick Jones and Erin Robinson to draft WSSSPE3 report back.

\item Before February 2016, Nick Jones and Erin Robinson to call a meeting of
the broader group, to review key next steps.
    
\item Second quarter 2016 -- if willing parties are identified, draft workshop proposal
and identify a relevant forum, including future WSSSPE events.
    
\end{enumerate}

\subsection{Description of Funding Needed}

Workshop/RCN travel funding to bring together key program, project, and funder
representatives from across North America, EU, UK, Australasia. In addition,
funding to support work on better defining the landscape of training activities,
the useful perspectives in communicating the value of the varied training
projects, and the possible pathways through training activities over time.

\section{Software Credit Working Group Discussion}
\label{sec:appendix_SW_credit}

Kyle Niemeyer\footnote{email:
\href{mailto:kyle.niemeyer@oregonstate.edu}{kyle.niemeyer@oregonstate.edu}} will
serve as the point of contact for this working group, and be responsible for
ensuring timely progress of the planned actions.

\subsection{Group Members}

\begin{itemize}
\item Alice Allen -- Astrophysics Source Code Library 
\item Sou-Cheng Choi -- NORC at University of Chicago, Illinois Institute of Technology
\item James Hetherington -- University College London
\item Lorraine Hwang -- University of California, Davis
\item Daniel S.\ Katz -- University of Chicago, Argonne National Laboratory
\item Frank L\"{o}ffler -- Louisiana State University
\item Abigail Cabunoc Mayes -- Mozilla Science Lab
\item Kyle E.\ Niemeyer -- Oregon State University
\item Grace Peng -- National Center for Atmospheric Research
\item Ilian Todorov -- Science \& Technology Facilities Council, UK
\end{itemize}

\subsection{Summary of Discussion}

The following section summarizes the working group's discussion based on
contributions prior to the meeting~\cite{WSSSPE3-SC-github-issues} and the
collaborative notes taken during the meeting~\cite{WSSSPE3-SC-google-notes}.
Please refer to the original sources for the unedited discussions if necessary.

Initial discussions focused on both various mechanisms for, and the
philosophical approach behind, crediting software in scientific papers. These
began with proposals for various ways to credit software (or other research
products including data) that contributed more significantly than a generic
citation, including:
\begin{itemize}

\item A hierarchy of citations, with a ``substantial'' citation category to
indicate software or data that played a more significant role in the research;

\item Transitive credit~\cite{wssspe2_katz,Katz:2014_tc}, which assigns
contriponents (contributors and components) various weights depending on their
level of importance;
    
\item Project CRediT~\cite{projectcredit}, which assigns roles to paper authors
based on their specific contributions; and
    
\item Mozilla Science Lab's recently introduced Contributorship Badges for
Science~\cite{Mozilla_badges}, which provide a badge---associated with an
ORCID~\cite{orcid}---that recognizes author contributions using the taxonomy
outlined in Project CRediT.
    
\end{itemize}
However, as of this writing, only Project CRediT
roles~\cite{McCall2015_credit,Lin2015_credit} and Contributorship
Badges~\cite{Mozilla_badges} have been implemented for published papers, and
both of these only provide a single ``Software'' or ``Computation'' category
associated with software. In addition, neither of these options allows for the
citation of software itself, but only provide an author contribution related to
software. The discussion quickly focused on transitive credit as a more
quantitative measure of allocating credit to both authors and software, although
there were some concerns about authors overestimating their own contributions
compared to prior work.

The discussion then evolved into philosophical questions about the importance or
reliance of a particular work on prior science, materials, or software---in
other words, whether there is a difference between depending on prior scientific
advances and depending on certain software (or experimental equipment).
Multiple contributors converged on the conclusion that unique capabilities
require some additional credit. The---albeit limited---consensus was that if a
particular study relied on the unique capabilities of software, data, or an
experimental apparatus, then the authors or developers that created this
capability should be credited somehow.

The group also agreed on the fact that additional data was required to support
the assertion that software was not being sufficiently cited in the literature.
In particular, this issue seemed to be field-dependent. For example, as shown by
a study of Howison and Bullard~\cite{Howison2015}, in the field of biology, the
most-cited papers appear to be those describing scientific software. However,
this may not---and likely is not---the case in other fields, nor is it clear
whether developers of scientific software, even in the case of the biology
field, are receiving sufficient credit for their efforts.

In the breakout sessions on the first day of WSSSPE3, the group discussed and deliberated over the
Entertainment Identifier Registry (EIDR)~\cite{EIDR} as a potential model for
scientific software. That system assigns unique Digital Object Identifiers
(DOIs)---the same system used for scientific publications---to all content
(e.g., movies, television shows) and contributors, along with relevant metadata.
One important use of the EIDR system is to track rights and credits for contributors
to entertainment works in order to distribute revenues---similar to the proposed
transitive credit concept.

The group also discussed separating quantitative measures (e.g., number of
citations) from the value of a work in order to give credit, moving towards
qualitative or anecdotal evidence of value. Other topics that were brought up included a
form of PageRank~\cite{Brin1998} for citations, based on number of mentions, and
using market penetration or adoption rate in a community as a metric, although
it was not clear how this would be measured. Finally, the concept a software
tool's uniqueness or indispensability to a community was mentioned, with value
being characterized by a particular piece of software either offering unique
capabilities or doing something better, faster, or with less computational
requirements than other offerings.

On the second day of WSSSPE3, the group decided to put aside the
taxonomy of contributions and focus on software citations to ensure developers
receive credit (regardless of contribution). Eventually, once 
software citations are standardized, the goal would be to return to establishing different
roles\slash contributions for this credit. Following this decision, the group
identified two necessary actions to move forward:
\begin{enumerate}

\item standardizing a citation file or some other form of metadata associated
with software, and

\item standardizing the way to cite software (used directly) in papers.
        
\end{enumerate}
For both of these actions, a number of ongoing efforts were identified and discussed.

\subsubsection{Software Citation Metadata}

At a minimum, the metadata required for software citation includes:
\begin{itemize}
    \item Name of software,
    \item DOI for software,
    \item Contributors, in the form of names and ORCIDs,
    \item Software dependencies, in the form of DOIs, and
    \item Other people and artifacts that would be cited or acknowledged in a paper.
\end{itemize}
This information would then be contained in a citation file, e.g., as part of
the GitHub repository. The group also discussed similar efforts such as
CodeMeta\footnote{CodeMeta: \url{https://github.com/codemeta/codemeta}}, an
attempt to codify minimal metadata schemes in JSON and XML for scientific
software and code, and implementing transitive credit via
JSON-LD~\cite{wssspe2_katz}. Some questions arose about how this information
would be stored for closed-source software.

As one mechanism for constructing accurate contributor lists from existing
project contributors, the group discussed associating GitHub accounts---as well
as accounts on Bitbucket, CodePlex, and other repositories for open-source
scientific software---with ORCID accounts. However, a (quick) response from
GitHub (via Arfon Smith) indicated that this might not be possible in the near
future: ``GitHub doesn't have any plans to allow ORCID accounts to be associated
with GitHub user accounts.''

\subsubsection{Citing Software in Publications}

Although far from a standard practice, examples of citing software in
publications can be found in various scientific communities---notably,
representative samples can be found in astronomy~\cite{astronomy_SW_examples}
and biology~\cite{Howison2015}. The group recommended collecting similar
examples from other communities, and then developing a software citation principles
document in concert with the FORCE11 Software Citation Working Group (see
\S\ref{SC:plan} for more details), following the model of the FORCE11 Data
Citation Principles document~\cite{DataCitation2014}.

The group further discussed briefly whether software used directly in a
publication---whether to perform simulation or analysis, or as a dependency for
newly developed software---should be distinguished from other references due to
the dependence of the study on these research artifacts. Suggestions included a
separate list of citations (with DOIs) for software and other research objects
that serve this sort of ``vital'' role. Similar recommendations were made by the
credit breakout group at WSSSPE2~\cite{WSSSPE2}.

Finally, although a discrete task from software citations,
significant discussion focused on ensuring software citations are indexed in the
same manner as publications, allowing the construction of a corresponding
software citation network. Currently, software releases can receive citable DOIs
via Zenodo~\cite{zenodo-web} and figshare~\cite{figshare-web}; however, these
citations are not processed by indexers such as Web of Science, Scopus, or
Google Scholar. Thus, either in parallel or following the primary task, the
group will need to reach out to these organizations. Initial conversations with
Elsevier\slash Scopus via Michael Taylor during WSSSPE3 clarified that Scopus
is not yet DataCite DOI aware, and also does not yet have an internal identifier
for software or data (but needs\slash plans to add this support). Taylor said
they prefer a ``software article'' with the usual article metadata (e.g.,
authors, citations), and mentioned Zenodo as an example -- this proposal seemed
to align with our group's discussions. Taylor also mentioned another benefit of the
software and associated DOI on GitHub: in addition to a citation, one could obtain
statistics on usage/downloads/forks, which happens to be what
Depsy\footnote{Depsy: \url{https://depsy.org}} is beginning to try to do.

\subsection{Description of Opportunity, Challenges, and Obstacles}

There currently is no standard mechanism for citing software or
receiving credit for software (akin to citations for publications). Software is
eligible for DOI assignment, but DOI metadata fields are not well tuned or
standardized for software (vs.\ publications). Some software providers apply for
DOIs, but this is not widely adopted. Also, there is no mechanism to cite
software dependencies within software.

Major obstacles include the fact that indexers (e.g., Scopus, Web of Science,
Google Scholar) do not currently support software/data document types or
DataCite DOIs. Therefore, even with universal association of scientific software
with DOIs and standardized practices for citing software in publications,
software citations will not be indexed in the same manner as traditional
publications.

Although this working group's discussions at WSSSPE3 did not focus much on the
topic of tenure and professional advancement, 
the group recognized that there is no standard policy---generally even within a
single university---for software products to be included in promotion and tenure
dossiers. Thus, it may be difficult to encourage valuing software contributions
across the United States or United Kingdom and globally; furthermore,
stakeholders are typically not tenured and thus may not be influential enough to
change the status quo. However, as discussed in Section~\ref{RSE}, this is
changing for Research Software Engineers, at least in the UK.

\subsection{Key Next Steps}
\label{SC:next-steps}

\begin{enumerate}

\item Hold virtual meeting to determine group members responsible\slash willing
to work on the following tasks, to be organized within one month of the workshop.

\item Compile best practices of software citation across multiple disciplines,
including journals and communities of interest\slash practice in the research
world, to begin by December 2015.

\item Compile examples of including other products in promotion and tenure
dossier, to begin by December 2015.

\item Draft the Software Citation Principles document (including citation
metadata file), by April 2016.

\item Publish\slash release the Software Citation Principles document, by August
2016.

\item Reach out to journals, publishers, teachers\slash educators, indexers, and
professional societies---likely through meetings with key groups, to begin by
September 2016.

\end{enumerate}

\subsection{Plan for Future Organization}
\label{SC:plan}

The WSSSPE breakout group plans to join efforts related to citing software with
the FORCE11 Software Citation Working Group (FORCE11-SCWG)\footnote{FORCE11
Software Citation Working Group,
\url{https://www.force11.org/group/software-citation-working-group}}; Kyle
Niemeyer formally requested the merging of these groups following the meeting.
However, some future plans of the WSSSPE group fall outside the scope of
FORCE11-SCWG, which covers software citation practices. These activities include
working with indexers such as Web of Science and Scopus to index software
citations archived on, e.g., Zenodo or figshare, and pursuing the development of
an open indexing service; such plans will be pursued either separately or
through the formation of follow-on FORCE11 working groups.

The group will primarily communicate electronically, with Kyle Niemeyer
responsible for ensuring regular progress.

\subsection{What Else is Needed?}

The near-term actions of the group, focused mainly on software citation, do not
require any additional resources. However, connections with publishers and
indexers will be needed to pursue related activities, although the FORCE11-SCWG
may satisfy this need; in addition, some members of the group already reached
out to relevant contacts. Funding may be needed to organize meetings or for
group members to attend relevant meetings, as discussed further below.

\subsection{Key Milestones and Responsible Parties}

Following the meeting, Kyle Niemeyer formally requested the merging of software
citation activities with FORCE11-SCWG. Within a month of the meeting, Niemeyer will
organize a virtual meeting of the group and manage the division of
responsibilities for compiling existing practices of software citation and
including software\slash products in promotion and tenure dossiers. Building off
of these efforts, the next major milestone is drafting the Software Citation
Principles document in collaboration with the SCWG, targeted for April 2016.
While the existing directors of the SCWG, Arfon Smith and Dan Katz, lead the
efforts of that group towards the Software Citation Principles document, Kyle
will help coordinate contributions from the WSSSPE group members.

\subsection{Description of Funding Needed}

Some funding would be useful to support primarily travel to conferences for
group meetings (e.g., FORCE2016\footnote{FORCE2016,
\url{https://www.force11.org/meetings/force2016}}), and to hold meetings to
bring together both group members and key stakeholders (e.g., journals,
publishers, professional societies, indexers). In addition, funding would be
desired to support group members' time to perform work towards the key steps
described previously.

\section{Publishing Software Working Group Discussion}
\label{sec:appendix_publishing_SW}

Steven R.\ Brandt\footnote{email:
\href{mailto:sbrandt@cct.lsu.edu}{sbrandt@cct.lsu.edu}} will serve as the point
of contact for this working group.

\subsection{Group Members}

\begin{itemize}
\item Steven R.\ Brandt -- Louisiana State University
\item Daniel Gunter -- LBNL
\item Yuhan Ding -- Illinois Institute of Technology
\item Neil Chue Hong -- Software Sustainability Institute
\end{itemize}

\subsection{Summary of Discussion}

A tentative first cut at the list containing executable papers identified the following:
\begin{itemize}

\item ACM Transactions on Mathematical Software (TOMS): provides the additional step
of having reviewers validate the code which was submitted with the publication.
 
\item The Mathematica Journal: publishes Mathematica notebooks (with equations,
figures, etc.) directly.

\item O'Reilly Media: announced that it plans to make IPython Notebooks a
first-class authoring environment for their publishing program alongside their
existing mechanisms.

\item Nature: offers a list of notebooks published alongside more traditional
articles, and is looking at ways to make these documents more official. There
are, in fact, a number of journals that offer ``electronic supplements'' to the
more traditionally published static articles.

\item IPython: maintained a list of ``reproducible academic
publications''~\cite{ipython-pubs}.

\item KBase: offers narratives built on IPython or Jupyter notebooks for assembling
publications that are reproducible, and can be commented or annotated.
  
\end{itemize}

The group also discussed future possibilities for a new publication format
that might provide advantages:
\begin{itemize}

\item Journals could be built around an existing, widely used framework thereby
reducing the burden of studying code on the part of reviewers (common bits of
infrastructure which are not relevant to the science would be automatically
excluded).

\item Journals might be encouraged to use more metadata, making them easier to
mine for various analytical purposes.

\item The Research Ideas and Outcomes (RIO) journal is an effort to publish
fragmentary results that can subsequently be combined into a single content
item.

\item Papers could be made more understandable. Each equation or technical term
could be linked to a document/tutorial explaining its origin and/or
derivation.

\item So many options for publication currently exist that good science may be
getting lost in the noise. Would some sort of ``upvote'' mechanism be of value?

\item Some sort of Replicated Computation Results badge could be made available
to publications that have undergone greater scrutiny (this is already done by
TOMS).
  
\end{itemize}

\subsection{Description of Opportunity, Challenges, and Obstacles}

The opportunity is to collect a list of executable papers and shine a
light on the experiments and development efforts currently underway.

The only obstacle to this is the difficulty in finding and identifying such
publications. The Software Sustainability Institute was able to do something
similar for publications about software by making a public page on the Software
Sustainability Institute's website (\url{http://www.software.ac.uk}) containing a
catalog of these publications and enlisting the help of the community to grow
the list. 

\subsection{Key Next Steps}

Create the first version of the web page to be displayed on the Software
Sustainability Institute's website: \url{http://www.software.ac.uk}.
We expect the page to be live in early January of 2016.

An ongoing effort to update the page should follow.

\subsection{Plan for Future Organization}

None at this time.

\subsection{What Else is Needed?}

Nothing else at this time.

\subsection{Key Milestones and Responsible Parties}

Steven R.\ Brandt has created a first version of the page, and it is in the process
of being posted on the Software
Sustainability Institute's website: \url{http://www.software.ac.uk}. 
Neil Chue Hong will take responsibility for the page once it is up.

\subsection{Description of Funding Needed}

None.

\section{User Community Working Group Discussion}
\label{sec:appendix_user_community}

Point of contact:
Dan Gunter\footnote{email: \href{mailto:dkgunter@lbl.gov}{dkgunter@lbl.gov}} and 
Ethan Davis\footnote{email: \href{mailto:edavis@ucar.edu}{edavis@ucar.edu}}.

\subsection{Group Members}

\begin{itemize}
\item Ethan Davis -- UCAR Unidata
\item Dan Gunter -- Lawrence Berkeley National Lab
\item Liz Jessup -- University of Colorado
\item Mark Miller -- University of California, San Diego
\item Lindsey Powers -- The HDF Group
\item Daniel Ziskin -- NCAR Atmospheric Chemistry Observations and Modeling (ACOM) Laboratory
\end{itemize}

\subsection{Summary of Discussion}

Discussion revolved around a few questions: what is the benefit of having a
``community'' for software sustainability, what practices and circumstances lead
to having and maintaining a community, how can funding help or hinder this
process, and perhaps most importantly, how can best practices be described and
distilled into a document that can help new projects.

The benefits of having a community that were brought up were considered largely
obvious. In addition to having advocates for the software, and a possible source
of ``free'' contributions to the codebase, the community becomes a good source
for requirements, feedback, and metrics. The software community can also act as
``cheerleaders'' who convince funders or other potential users to fund/use the
software, and thus help sustain the software.

Practices and circumstances that lead to a community are first, that the
software offers value. But in addition to this, a community will be much more
likely to form if they receive (expert) support when they have questions.
Additional contributing factors are good usability (not always needed), and an
open development process such as IPython developer meetings on YouTube. It was
also pointed out that an evangelist for the project, not necessarily but often
one of the developers, can often make a big difference.

Funding can help the process by encouraging both value to the community and
high-quality user support. Only providing funding for the software development
may create good software, but with less likelihood to have a real community. It
was discussed that federal laboratories are a good incubator for software
communities, and that a general facility like EarthCube is too dispersed to
really make a community. Also, domain-specific groups within laboratories or
universities might provide as an incubator for software communities.

In describing best practices, the group discussed the different modes for
starting a scientific software project: building on an existing product that
needs improving, recognizing an unsatisfied need of an existing community, or
creating a new solution to a need not yet recognized by the community. The group
also thought that the existing books on software communities would need to be
evaluated in light of differences between science software projects and general
open-source software (OSS) projects in terms of scale, science, acknowledgement and credit, and funding
models.

\subsection{Description of Opportunity, Challenges, and Obstacles}

The main opportunity is to increase awareness among scientific
software developers and project managers of the importance of
developing a community around their project.
While this message is fairly well understood in the open source
community, the scientific community can be more focused on the
science a software project is supporting rather than the software
project itself.

As with many of the issues relevant to the sustainability of science
software, the main challenge here will be changing the culture and
expectations around scientific software.

\subsection{Key Next Steps}

The most important next steps is a ``Best Practice'' document, which would
describe what successful projects with engaged communities look like, how to
replicate this type of project, and look at end-of-life on a community project.
Inputs to this document would include a software community survey of highly
functioning communities such as R Open Science, Python SciPy, OPeNDAP, and
Unidata, with analysis of factors that feed into their success. Also references
like the ``Art of Community'' could be adapted and summarized for the science
software community.

More specifically, the group would like to take the following steps:

\begin{itemize}
\item Survey successful science software projects
\item Survey community members from the surveyed projects
\item Distill the survey results and document best practices around community engagement
\item Look for ways to raise awareness
\end{itemize}

Another next step would be increasing recognition of need for science software
projects to focus on building and supporting their user communities. Good software
engineering practices are not enough, and popular training like Software
Carpentry does not currently address this issue head-on.

\subsection{Plan for Future Organization}

No definite plans were agreed upon for future organization. The major ideas discussed
were coordinating with another group or adapting some existing text.

Collaboration within the framework of an existing organization seems a good initial
path. Mozilla Science maintains a ``Working Open Project
Guide''~\cite{working-open-wssspe3}, the introduction of which states:
\begin{quote}
Working openly with contributors enables your community to learn how to build
and collaborate together. This document is a guideline on how to work openly and
involve others in your projects with Mozilla. We want to help you engage your
community in a way that encourages contributors and builds other leaders.
 \end{quote}

 Another idea is to form a group that could adapt existing commercial-oriented 
 guidelines for the world of scientific software and top-down funding structures.
 For example, to distill the ``Art of Community'' by Jono Bacon~\cite{art-of-community}
for scientific software.

\subsection{What Else is Needed?}

The group had many points of agreement, but there is not currently a dedicated core group
of people who have committed to producing the key milestones. Coordination via phone or
online would be necessary to build this ``community'' of contributors.

\subsection{Key Milestones and Responsible Parties}

The key milestones for the group's activities align closely with the Key Next Steps above:

\begin{itemize}
\item Complete and write up a survey of project members, and community members, for successful science software projects
\item Distill the survey results and document best practices around community engagement
\end{itemize}

\subsection{Description of Funding Needed}

With a small amount of seed funding, it is possible that members of this group or other parties could
spend the time necessary to devise a survey of existing projects and deploy this, probably by traveling to
meetings and workshops for the various software communities.

\bibliographystyle{vancouver}

\bibliography{wssspe}

\begin{thebibliography}{10}

\bibitem{WSSSPE1-pre-report}
Katz DS, Allen G, {Chue Hong} N, Parashar M, Proctor D.
\newblock First Workshop on on Sustainable Software for Science: Practice and
  Experiences {(WSSSPE)}: Submission and Peer-Review Process, and Results.
\newblock arXiv; 2013. 1311.3523.
\newblock \url{http://arxiv.org/abs/1311.3523}.

\bibitem{WSSSPE1}
Katz DS, Choi SCT, Lapp H, Maheshwari K, L\"{o}ffler F, Turk M, et~al.
\newblock Summary of the First Workshop on Sustainable Software for Science:
  Practice and Experiences ({WSSSPE1}).
\newblock Journal of Open Research Software. 2014;2(1).
\newblock \url{http://dx.doi.org/10.5334/jors.an}.

\bibitem{WSSSPE2-pre-report}
Katz DS, Allen G, {Chue Hong} N, Cranston K, Parashar M, Proctor D, et~al.
\newblock Second Workshop on Sustainable Software for Science: Practice and
  Experiences ({WSSSPE2}): Submission, Peer-Review and Sorting Process, and
  Results.
\newblock arXiv; 2014. 1411.3464.
\newblock \url{http://arxiv.org/abs/1411.3464}.

\bibitem{WSSSPE2}
Katz DS, Choi SCT, Wilkins-Diehr N, {Chue Hong} N, Venters CC, Howison J,
  et~al.
\newblock Report on the Second Workshop on Sustainable Software for Science:
  Practice and Experiences {(WSSSPE2)}.
\newblock Journal of Open Research Software. 2016;Accepted. Available at
  \url{http://arxiv.org/abs/1507.01715}.

\bibitem{2011ApJS..192....9T}
{Turk} MJ, {Smith} BD, {Oishi} JS, {Skory} S, {Skillman} SW, {Abel} T, et~al.
\newblock {yt: A Multi-code Analysis Toolkit for Astrophysical Simulation
  Data}.
\newblock ApJS. 2011 Jan;192:9.

\bibitem{Becker:2014}
Becker C, Chitchyan R, Duboc L, Easterbrook S, Mahaux M, Penzenstadler B,
  et~al.
\newblock {The Karlskrona manifesto for sustainability design}.
\newblock arXiv; 2014. 1410.6968.
\newblock \url{http://arxiv.org/abs/1410.6968}.

\bibitem{sempervirens}
{Sempervirens};.
\newblock Accessed: 2015-11-07.
\newblock \url{https://github.com/njsmith/sempervirens}.

\bibitem{5069157}
Heroux MA, Willenbring JM.
\newblock Barely sufficient software engineering: 10 practices to improve your
  CSE software.
\newblock In: Software Engineering for Computational Science and Engineering,
  2009. SECSE '09. ICSE Workshop on; 2009. p. 15--21.

\bibitem{Blatt_WSSSPE}
Blatt M.
\newblock {DUNE} as an Example of Sustainable Open Source Scientific Software
  Development.
\newblock arXiv; 2013. 1309.1783.
\newblock \url{http://arxiv.org/abs/1309.1783}.

\bibitem{Ahern_WSSSPE}
Ahern S, Brugger E, Whitlock B, Meredith JS, Biagas K, Miller MC, et~al.
\newblock {VisIt}: Experiences with Sustainable Software.
\newblock arXiv; 2013. 1309.1796.
\newblock \url{http://arxiv.org/abs/1309.1796}.

\bibitem{Vliet:2008}
Vliet Hv.
\newblock Software Engineering: Principles and Practice.
\newblock 3rd ed. Wiley Publishing; 2008.

\bibitem{Merali:2010}
Merali Z.
\newblock {Computational science:...Error..why scientific programming does not
  compute}.
\newblock Nature. 2010;467:775--777.

\bibitem{hettrick:2014}
Hettrick S, et~al.. UK Research Software Survey 2014; 2014.
\newblock Available from: \url{http://dx.doi.org/10.5281/zenodo.14809}.

\bibitem{Becker:2016}
Becker C, Betz S, Chitchyan R, Duboc L, Easterbrook SM, Penzenstadler B, et~al.
\newblock Requirements: The Key to Sustainability.
\newblock Software, IEEE. 2016 Jan;33(1):56--65.

\bibitem{Becker:2015}
Becker C, Chitchyan R, Duboc L, Easterbrook S, Penzenstadler B, Seyff N, et~al.
\newblock {Sustainability Design and Software: The Karlskrona Manifesto}.
\newblock In: {Proc. 2015 {Int'l Conf. Software Eng. (ICSE'15)},}; 2015. .

\bibitem{tate:2005}
Tate K.
\newblock Sustainable Software Development: An Agile Perspective.
\newblock Addison-Wesley Professional; 2005.

\bibitem{swebokv3}
Pierre~Bourque REF.
\newblock SWEBOK, version 3.0: Guide to the Software Engineering Body of
  Knowledge.
\newblock IEEE Press; 2014.

\bibitem{WSSSPE}
{Working towards Sustainable Software for Science: Practice and Experiences};.
\newblock Accessed: 2015-12-03.
\newblock \url{http://wssspe.researchcomputing.org.uk/}.

\bibitem{SEHPCCSE}
{International Workshop on Software Engineering for High Performance Computing
  in Computational Science and Engineering};.
\newblock Accessed: 2015-12-03.
\newblock \url{http://se4science.org/workshops/}.

\bibitem{se4susy}
{Workshop on Software Engineering for Sustainable Systems};.
\newblock Accessed: 2015-12-03.
\newblock \url{http://sustainabilitydesign.org/initiatives/se4susy/}.

\bibitem{EIDR}
{Entertainment Identifier Registry};.
\newblock Accessed: 2015-10-28.
\newblock \url{http://eidr.org}.

\bibitem{wssspe2_katz}
Katz DS, Smith AM.
\newblock Implementing Transitive Credit with {JSON-LD}.
\newblock arXiv; 2014. 1407.5117.
\newblock \url{http://arxiv.org/abs/1407.5117}.

\bibitem{Katz:2014_tc}
Katz DS.
\newblock Transitive Credit as a Means to Address Social and Technological
  Concerns Stemming from Citation and Attribution of Digital Products.
\newblock Journal of Open Research Software. 2014 Sep;2(1):e20.

\bibitem{working-open-wssspe3}
Mayes AC, zee moz, Collins A, Niemeyer K, Jabbari A. {Leadership-Training:
  ``Working Open'' Guide - WSSSPE3 version}; 2015.
\newblock Available from: \url{http://dx.doi.org/10.5281/zenodo.33748}.

\bibitem{art-of-community}
Bacon J.
\newblock The Art of Community. Building the New Age of Participation.; 2009.

\bibitem{howison_sustaining_2015}
Howison J.
\newblock Sustaining scientific infrastructures: transitioning from grants to
  peer production (work-in-progress).
\newblock In: {iConference} 2015 Proceedings; 2015.
  \url{http://hdl.handle.net/2142/73439}.

\bibitem{Geels:2007}
Geels FW, Schot J.
\newblock Typology of sociotechnical transition pathways.
\newblock Research Policy. 2007;36(3):399--417.

\bibitem{WSSSPE3-SC-github-issues}
{{WSSSPE3} Software Credit Working Group}. {WSSSPE3} Software Credit Working
  Group {GitHub} Issues; 2015.
\newblock Accessed: 2015-10-1.
\newblock \url{https://github.com/WSSSPE/meetings/issues/51}.

\bibitem{WSSSPE3-SC-google-notes}
{WSSSPE3 Software Credit Working Group}. {WSSSPE3} Software Credit Working
  Group Collaborative Notes; 2015.
\newblock Accessed: 2015-10-1.
\newblock
  \url{https://docs.google.com/document/d/1oN0ZYqIoWtOE1LBMIlWY9N8nn5LHTncj8GjUKPh62pA}.

\bibitem{projectcredit}
CASRAI. Project {C}redit;.
\newblock Accessed: 2015-03-31.
\newblock \url{http://credit.casrai.org}.

\bibitem{Mozilla_badges}
Mayes AC. Contributorship Badges; 2015.
\newblock Accessed: 2015-10-26.
\newblock \url{https://www.mozillascience.org/projects/contributorship-badges}.

\bibitem{orcid}
Open {R}esearcher and {C}ontributor {ID} ({ORCID});.
\newblock Accessed: 2015-03-31.
\newblock \url{http://orcid.org/}.

\bibitem{McCall2015_credit}
McCall JG, Al-Hasani R, Siuda ER, Hong DY, Norris AJ, Ford CP, et~al.
\newblock {CRH} Engagement of the Locus Coeruleus Noradrenergic System Mediates
  Stress-Induced Anxiety.
\newblock Neuron. 2015 Aug;87(3):605--620.

\bibitem{Lin2015_credit}
Lin IC, Okun M, Carandini M, Harris KD.
\newblock The Nature of Shared Cortical Variability.
\newblock Neuron. 2015 Aug;87(3):644--656.

\bibitem{Howison2015}
Howison J, Bullard J.
\newblock Software in the scientific literature: {Problems} with seeing,
  finding, and using software mentioned in the biology literature.
\newblock Journal of the Association for Information Science and Technology.
  2015;In press, available at \url{http://dx.doi.org/10.1002/asi.23538}.

\bibitem{Brin1998}
Brin S, Page L.
\newblock The anatomy of a large-scale hypertextual Web search engine.
\newblock Computer Networks and {ISDN} Systems. 1998;30(1--7):107--117.

\bibitem{astronomy_SW_examples}
{SIG} SP. Astronomy software citation examples and ideas; 2015.
\newblock Accessed: 2015-10-29.
\newblock
  \url{https://docs.google.com/document/d/1q9ULl7alA3veL7Qwg7jGteRWeJwlrkvRHSXjvt-rTs0}.

\bibitem{DataCitation2014}
{Data Citation Synthesis Group}. Martone M, editor. Joint Declaration of Data
  Citation Principles.
\newblock San Diego CA: FORCE11; 2014.
\newblock Accessed: 2015-10-28.
\newblock
  \url{https://www.force11.org/group/joint-declaration-data-citation-principles-final}.

\bibitem{zenodo-web}
{Zenodo};.
\newblock Accessed: 2015-10-28.
\newblock \url{https://zenodo.org}.

\bibitem{figshare-web}
{figshare};.
\newblock Accessed: 2014-02-03.
\newblock \url{https://figshare.com}.

\bibitem{ipython-pubs}
{IPython}. A gallery of interesting {IPython Notebooks}; 2015.
\newblock Accessed: 2015-10-28.
\newblock
  \url{https://github.com/ipython/ipython/wiki/A-gallery-of-interesting-IPython-Notebooks#reproducible-academic-publications}.

\end{thebibliography}
\end{document}